\documentclass{sn-jnl}

\usepackage{stmaryrd}
\usepackage{enumitem}
\usepackage{graphicx}
\usepackage{amsmath}
\usepackage{amssymb}
\usepackage{amsfonts}
\usepackage{dsfont}
\usepackage{mathtools}
\usepackage{mathrsfs}
\usepackage{algorithm}
\usepackage{algpseudocode}
\usepackage{tcolorbox}
\usepackage{float}
\floatstyle{plain}
\restylefloat{table}
\usepackage[numbers]{natbib}
\newtheorem{assumption}{Assumption}

\newtheorem{example}{Example}

\newcommand{\indep}{\mathrel{\perp\!\!\!\perp}}

\begin{document}

\title{A General Framework for Joint Multi-State Models}

\author[1]{\fnm{F\'elix} \sur{Laplante}} \email{felixlaplante.research@gmail.com}
\author*[2]{\fnm{Christophe} \sur{Ambroise}} \email{christophe.ambroise@univ-evry.fr}

\affil*[1]{\orgdiv{MaIAGE}, \orgname{Universit\'e Paris-Saclay, INRAE},
    \orgaddress{\country{France}}}
\affil[2]{\orgdiv{Laboratoire de Math\'ematiques et Mod\'elisation d'\'Evry (LaMME)},
    \orgname{Universit\'e Paris-Saclay, CNRS, Univ \'Evry},
    \orgaddress{\country{France}}}

\abstract{
    Conventional joint modeling approaches generally characterize the relationship between longitudinal biomarkers and discrete event occurrences within terminal, recurring or competing risk settings, thereby offering a limited representation of complex, multi-state trajectories.

    We propose a general multi-state joint modeling framework that unifies longitudinal biomarker dynamics with multi-state time-to-event processes defined on arbitrary directed graphs. The proposed framework also accommodates nonlinear longitudinal submodels and scalable inference via stochastic gradient ascent. This formulation encompasses both Markovian and semi-Markovian transition structures, allowing recurrent cycles and terminal absorptions to be naturally represented. The longitudinal and event processes are linked through shared latent structures within nonlinear mixed-effects models, extending classical joint modeling formulations.

    We derive the complete likelihood, model selection criteria, and develop scalable inference procedures based on stochastic gradient ascent to enable high-dimensional and large-scale applications. In addition, we formulate a dynamic prediction framework that provides individualized state-transition probabilities and personalized risk assessments along complex event trajectories.

    Through simulation and application to the PAQUID cohort, we demonstrate accurate parameter recovery and individualized prediction.
}


\maketitle

\section{Introduction} \label{sec:introduction}

Joint modeling of longitudinal and time-to-event data has become an essential tool of modern biostatistics \citep{papageorgiou2019overview}, particularly for dynamic prediction in clinical applications. Classical joint models typically couple a longitudinal biomarker process with a single time-to-event outcome \citep{rizopoulos2012joint}, allowing for the integration of biological knowledge and individual heterogeneity via shared latent structures. However, many real-world processes involve multiple possible outcomes, intermediate stages, or recurrent events, which cannot be fully captured by a single-event framework \citep{KrolMauguen2017}. To address these limitations, specialized frameworks have been developed to handle competing risks and recurrent events. However, these remain constrained to specific event types and fail to accommodate more general multi-state transitions. In such settings, multi-state models provide a more general and flexible approach \citep{ferrer2016joint, you2024joint}.

Multi-state models represent an overall joint process of clinical course as a series of discrete stages or health states that occur sequentially \citep{lovblom2024modeling}. In biostatistics, these models are widely used for survival and reliability analysis, allowing for a richer and more accurate representation by capturing alternative paths to an event of interest, intermediate events, and progressive disease. Key components of multi-state models include transition intensity functions, which denote the instantaneous risk of moving from one state to another, and transition probability functions, which describe the probability of transition over longer intervals. Often, these models assume the Markov property, where future transitions depend only on the current state, simplifying the transition intensity functions \citep{asanjarani2022estimation}. This assumption can be relaxed by adopting a semi-Markov formulation, in which the transition probabilities depend not only on the current state but also on the sojourn time, effectively resetting the time scale after each transition.

The link between multi-state models and joint models arises when a multi-state process is integrated as a component within a broader joint modeling framework. While multi-state models, such as those described in the ``sequential state framework'' primarily focus on movements between discrete states, joint models operate under a ``parallel trajectory framework'' that combines a longitudinal process with a time-to-event process.

The model proposed by Ferrer et al.~\citep{ferrer2016joint} exemplifies this link by presenting a joint model for a longitudinal process (e.g., Prostate-Specific Antigen (PSA) measurements) and a multi-state process (e.g., clinical progressions in prostate cancer). These two sub-models are interconnected by shared random effects, allowing the model to account for the correlation between a continuous longitudinal biomarker trajectory and acyclic discrete transitions between health states. Specifically, the previously proposed mechanism \citep{ferrer2016joint, you2024joint} is based on binary censoring indicators for each state, implying that any given state can be entered at most once, therefore precluding the modeling of recurrent events.

The model we propose operates on an arbitrary directed graph, enabling representation of complex state transitions and recurrent events. We derive the complete likelihood for the nonlinear joint model, introduce an efficient stochastic approximation inference method, and develop dynamic prediction tools. To illustrate its practical relevance, we conduct both a simulation study and a synthetic biomedical case study.

This paper is organized as follows. Section~\ref{sec:background} reviews the background on multi-state and joint modeling frameworks. Section~\ref{sec:model} introduces the proposed general multi-state joint modeling framework, detailing the likelihood formulation and proposed model selection criteria. Section~\ref{sec:inference} describes the stochastic ascent inference algorithm, while Section~\ref{sec:prediction} presents the dynamic prediction framework. Section~\ref{sec:experiments} reports simulation experiments that assess parameter recovery and convergence properties of the proposed model, whereas Section~\ref{sec:paquid} applies the framework to data from the PAQUID cohort \citep{Letenneur1994}. Finally, Section~\ref{sec:conclusion} concludes with perspectives and future research directions.

\section{Background} \label{sec:background}

Joint modeling of longitudinal and time-to-event data provides a unified statistical framework to analyze the interplay between continuous biomarker dynamics and event occurrence processes. In its classical form, this framework couples a mixed-effects model describing individual biomarker trajectories with a survival model for event times \citep{rizopoulos2012joint}. Such models have become central in biomedical research, particularly for dynamic prediction and personalized risk assessment \citep{papageorgiou2019overview}. However, these traditional joint models generally focus either on a single terminal event, competing, or recurrent events, limiting their ability to represent more complex event histories that include intermediate, recurrent, or competing events. To address these limitations, multi-state extensions have been developed, providing a natural framework for describing transitions between multiple clinical or biological states over time \citep{ferrer2016joint, KrolMauguen2017}.

\subsection{Multi-State Markov Processes}

A \emph{multi-state stochastic process} is defined as a process $S(t)$ for $t \geq 0$, where $S(t)$ can take a finite number of values (states), often labeled $1, 2, \dots, p$. Quantities of interest typically include the probability of being in a certain state at a given time and the distribution of first passage times.

A \emph{Markov process} (or continuous-time Markov chain) is a
specific class of multi-state models in which future transitions between states \emph{depend only upon the current state}. This \emph{Markov property} means the process is \emph{memoryless}. A key consequence is that the duration spent in any state follows an \emph{exponential distribution}, implying a constant hazard rate for leaving that state \citep{jackson2011multi,putter2007competing}.

However, the Markov assumption can be unrealistic in many real-world applications. For example, in the study of human sleep stages, sojourn times often do not follow an exponential distribution \citep{dong2008non, roever2010statistical}, and in the case of chronic diseases such as AIDS, the risk of disease progression can depend on the time elapsed since infection \citep{andersen1999multi}. To address these limitations, \emph{semi-Markov processes} (SMPs) were introduced, which allow for arbitrary distributions of sojourn times while retaining the Markov property for the embedded discrete-time chain. This flexibility makes SMPs suitable for modeling complex disease progression and patient recovery scenarios \citep{putter2007competing}.

\subsection{Multi-State Semi-Markov Processes}

\emph{Multi-state semi-Markov processes} (MSMPs) offer a natural generalization by allowing the distribution of sojourn times to be arbitrary.

In an MSMP, the process is defined by a directed graph $G = (V, E)$,
where $V$ represents the set of states and $E$ the set of possible transitions. The transition intensities $\lambda_{k \to k'}(t \mid t_0)$ from state $k$ to state $k'$ at time $t$, given entry time $t_0$, are therefore assumed to be non-identically zero. The sojourn time in state $k$ is not restricted to an exponential distribution, allowing for more realistic modeling of the time spent in each state.

MSMPs have been widely applied in a range of disciplines. In reliability, they are used to model degradation and repair processes \citep{limnios2001semi}. In biomedical studies, they have proven useful for analyzing illness-death models or disease progression with non-exponential transitions \citep{commenges2006likelihood, fiocco2008modelling}. Applications in finance also exist, where semi-Markovian dynamics can model credit rating migrations or economic regimes \citep{luciano2006semi}. In these contexts, MSMPs retain the Markov property in the embedded jump chain while offering increased realism through flexible dwell time modeling.

From a methodological standpoint, MSMPs extend estimation strategies developed for Markov models, such as maximum likelihood or Bayesian inference, and can accommodate interval-censored or misclassified data \citep{frydman2005estimation}. This makes them a powerful and general tool for multi-state event history analysis.

While both joint models and multi-state models offer powerful tools, they primarily address different facets of disease progression and have complementary strengths. Multi-state models excel at understanding the sequence and timing of discrete events, while joint models are adept at modeling continuous biomarker trajectories and their associations with event outcomes for dynamic prediction. The complexity of real-world diseases often necessitates a more integrated approach that can leverage the strengths of both frameworks. For instance, in prostate cancer, multiple types of relapse may occur successively (a multi-state process), and their risk is influenced by the dynamics of longitudinal biomarkers like PSA. Previous research has initiated this integration by linking longitudinal data with a multi-state process \citep{ferrer2016joint}, without allowing the modeling of recurrent events.

\section{A General Multi-State Joint Modeling Framework}

To address the limitations imposed by traditional joint modeling frameworks, we extend them to incorporate a multi-state process, thereby enabling a unified analysis of longitudinal biomarkers and discrete transitions between multiple health states.

The proposed model defines a joint process $\{Y_i(t), S_i(t)\}_t$, where $Y_i(t)$ denotes the longitudinal biomarker trajectory and $S_i(t)$ the multi-state event process. Both components are linked through shared latent structures, and the event process is represented on an arbitrary directed graph supporting both Markovian and semi-Markovian transition mechanisms.

\subsection{Notation Conventions}

Throughout this section, let $i \in \llbracket n \rrbracket$ index $n \geq 1$ individuals, and let $t_{ij}$ denote the $j$-th observation time of individual $i$. The vector $Y_{ij} \in \mathbb{R}^d$ represents longitudinal biomarker measurements, and $X_i \in \mathbb{R}^p$ the associated covariates.  

The latent random effects $b_i \sim \mathcal{N}(0, Q)$ capture individual heterogeneity, $\gamma$ represent shared characteristics across all individuals, and the subject-specific parameters are defined as $\psi_i \coloneqq f(\gamma, X_i, b_i)$. The set $\mathcal{H}_i(t) \coloneqq \left\{ (s, h(s, \psi_i)) : s \leq t \right\}$ represents the true underlying history of the longitudinal process up to time $t$. Correspondingly, $\mathcal{Y}_i(t)$ denotes the observed noisy longitudinal history, and $\mathcal{S}_i(t)$ denotes the observed trajectory of the semi-Markovian dynamics.  

Transition times and states are denoted by $(T_{i\ell}, S_{i\ell}) \in \mathbb{R}^+ \times V$, where $V$ is the space of admissible states, and are subject to right-censoring at time $C_i \in \bar{\mathbb{R}}^+$. We define the index of the last observed transition before time $t$ as $m_i(t) \coloneqq \sup \{ \ell \geq 0 : T_{i\ell} \leq t \}$. The complete sequence of transition times and states for individual $i$ is denoted by $\mathcal{T}_i^*$, whereas $\mathcal{T}_i$ denotes the corresponding right-censored sequence.

\subsection{Multi-State Processes on Arbitrary Graphs} \label{sec:model}

The joint model proposed by Ferrer et al. \citep{ferrer2016joint} established a foundational link between longitudinal biomarkers and multi-state processes through shared random effects. However, their formalism relies on binary censoring indicators to characterize transitions, thereby inherently restricting the modeling framework to Directed Acyclic Graphs (DAGs) and excluding recurrent events. Indeed, their model uses, for each state $k \in V$, a tuple $(T_k, \delta_k)$, where $T_k$ denotes the observed or censored time until leaving state $k$ and $\delta_k$ is an indicator equal to $1$ if the transition was observed and $0$ if the observation was censored (\textit{i.e.}, the transition was not observed).

To overcome this limitation, we present a new method based on arbitrary graph transitions. By shifting away from binary indicators toward a more flexible graph-theoretic specification, our approach accommodates complex disease trajectories, including recurrent events and cyclic transitions that cannot be captured within a DAG-based structure.

\subsubsection{Graph Structure}

\begin{figure}[H]
    \centering
    \includegraphics[width=0.8\textwidth]{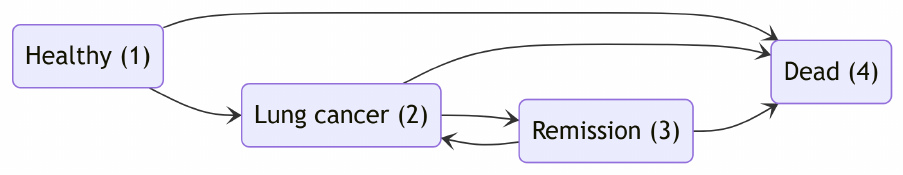}
    \caption{Example of a 4-state transition graph $G = (V,E)$, including
        an absorbing state (4).}
    \label{fig:state-graph}
\end{figure}

Let $G = (V,E)$ be a directed graph, where $V = \llbracket p \rrbracket$ denotes the set of states and $E \subseteq V \times V$ the set of admissible transitions. The graph encodes all possible paths of the multi-state process, allowing for competing, recurrent, or absorbing transitions. Figure~\ref{fig:state-graph} illustrates an example of a four-state model, including an absorbing state (Death). The corresponding adjacency matrix $A = (A_{kk'})_{1 \leq k, k' \leq p}$ satisfies $A_{kk'} = 1$ if $(k, k') \in E$ and $0$
otherwise.

\begin{figure}[H]
    \centering
    \begin{minipage}{0.31\textwidth}
        \centering
        \textbf{Standard Risk}
        \vspace{0.5em}
        \includegraphics[width=0.85\textwidth]{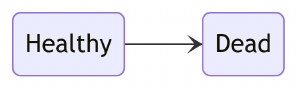}
    \end{minipage}
    \hfill
    \begin{minipage}{0.31\textwidth}
        \centering
        \textbf{Competing Risks}
        \vspace{0.5em}
        \includegraphics[width=0.9\textwidth]{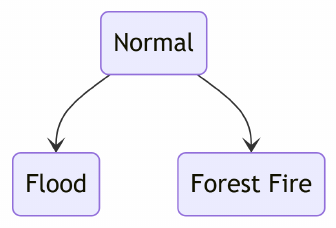}
    \end{minipage}
    \hfill
    \begin{minipage}{0.31\textwidth}
        \centering
        \textbf{Recurrent Risk}
        \vspace{0.5em}
        \includegraphics[width=\textwidth]{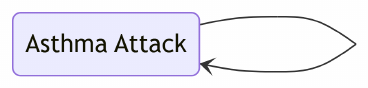}
    \end{minipage}
    \caption{Illustration of classical joint modeling approaches: (left) standard risk, (middle) competing risks, and (right) recurrent risk.\label{fig:joint-models}}
\end{figure}

Such a representation generalizes standard joint models, which can be recovered as special cases: single-event (linear chain), competing-risks (two absorbing transitions), or recurrent-event settings (cyclic transitions) as illustrated in Figure~\ref{fig:joint-models}.

Such joint modeling offers a flexible framework that extends beyond classical approaches and is particularly useful in medical applications.

\subsubsection{Individual Trajectories}

Each individual $i \in \llbracket n \rrbracket$ follows a latent trajectory
\[
    \mathcal{T}_i^* \coloneqq \left( (T_{i0}, S_{i0}), (T_{i1}, S_{i1}), (T_{i2}, S_{i2}), \dots \right),
\]
where $T_{i\ell} \in \mathbb{R}^+$ denotes the $\ell$-th transition time and
$S_{i\ell} \in V$ the corresponding state. The observed trajectory is right-censored at time $C_i \in \bar{\mathbb{R}}^+$, so that only $\mathcal{T}_i \coloneqq \left\{ (T_{i\ell}, S_{i\ell}) : T_{i\ell} \leq C_i \right\}$ is observed. Between transitions, the state $S_i(t)$ is constant, thus the transitions are the discontinuity points of the map $t \mapsto S_i(t)$, of which we assume there are at most countably many.

Letting $m_i(t) \coloneqq \sup \{ \ell \geq 0 : T_{i\ell} \leq t \}$ be the index of the last observed transition before time $t$ and $m_i \coloneqq m_i(C_i)$, we thus assume that no further transitions occur for individual $i$ on the interval $(T_{im_i}, C_i]$. We also define $\mathcal{S}_i(t) \coloneqq \left\{ (s, S_i(s)) : s \leq t \right\}$ as the observed trajectory up to time $t$, which is equivalent to specifying all transition times and their corresponding states up to $t$, together with the time $t$ itself.

The process is assumed to satisfy a (possibly time-inhomogeneous) semi-Markov property
\begin{equation} \label{ass:semi-markov}
    \mathscr{L} \left( (T_{i(\ell+1)}, S_{i(\ell+1)}) \mid \{(T_{i\ell'}, S_{i\ell'})\}_{\ell' \leq \ell} \right) = \mathscr{L} \left( (T_{i(\ell+1)}, S_{i(\ell+1)}) \mid (T_{i\ell}, S_{i\ell}) \right),
\end{equation}
which is relaxed compared to a stricter Markov property by allowing sojourn-time dependence.

\bigbreak

We emphasize once more that modeling the observed data solely as censored collections of transition times, as in Ferrer et al.~\citep{ferrer2016joint}, is not sufficient to represent such trajectories, since it implicitly restricts each state to be visited at most once and therefore rules out recurrent transitions.

\subsubsection{Transition Intensities and Survival Functions}

Although one could equivalently define intensities for all ordered pairs of states and set those corresponding to inadmissible transitions identically to zero, explicitly restricting attention to a graph of admissible transitions provides a clearer and more coherent representation of the underlying state dynamics.

\bigbreak

For any admissible transition $(k, k') \in E$, let $\lambda_i^{k \to k'}(t_0, t)$ denote the instantaneous risk (\textit{i.e.} \emph{hazard}) at time $t$ of moving to state $k'$, given entry time into state $k$ of $t_0 \leq t$
\[
    \lambda_i^{k \to k'}(t_0, t) \coloneqq \lim_{\delta \to 0^+}
    \frac{\mathbb{P}(T_{i(\ell+1)} \leq t + \delta, S_{i(\ell+1)}=k' \mid T_{i(\ell+1)} > t, T_{i\ell}=t_0, S_{i\ell}=k)}{\delta}.
\]

Assuming this quantity to be finite for any transition $(k, k') \in E$ as well as any pair of times $t \geq t_0$, we ensure that the underlying distribution of transition times is absolutely continuous with respect to the Lebesgue measure and admits a density. In this case, the cumulative risk and the corresponding survival function are
\begin{align}
     &\Lambda_i^{k \to k'}(t_0, t) = \int_{t_0}^{t} \lambda_i^{k \to k'}(t_0, s) \, ds, \notag \\
     &\mathbb{P}\left( T_{i(\ell+1)} > t \mid T_{i\ell}=t_0, S_{i\ell}=k \right) = \exp\left( -\sum_{j : (k, j) \in E}
        \Lambda_i^{k \to j}(t_0, t) \right) \label{eq:survival}.
\end{align}

The conditional transition probability is therefore
\begin{equation}
    \mathbb{P}\left( S_{i(\ell+1)} = k' \mid T_{i(\ell+1)}=t, T_{i\ell}=t_0, S_{i\ell}=k \right) = \frac{\lambda_i^{k \to k'}(t_0, t)}{\sum_{j : (k, j) \in E} \lambda_i^{k \to j}(t_0, t)} \label{eq:conditional},
\end{equation}
so for any $t \geq T_{i\ell}$, the joint density  writes as
\begin{align}
p\left( T_{i(\ell+1)}=t, S_{i(\ell+1)}=k' \mid T_{i\ell}=t_0, S_{i\ell}=k \right) = &\lambda_i^{k \to k'}(t_0, t) \notag \\
&\exp\left( -\sum_{j : (S_{i\ell}, j) \in E} \Lambda_i^{k \to j}(t_0, t) \right). \label{eq:transition-density}
\end{align}

The proofs of these various well-known formulae are postponed to the Appendix (see Section~\ref{sec:proof-risk}). In practice, numerical integration can be efficiently carried out using quadrature methods such as the Gauss-Legendre quadrature \citep{lether1978construction}.

\subsection{Multi-State Joint Model Specification}

Omitting the implicit conditioning on the individual parameters and covariates for the sake of simple notation, the proposed multi-state joint model with a Gaussian prior and homoscedastic Gaussian noise is specified by
\[
\begin{cases}
    Y_{ij} = h(t_{ij}, \psi_i) + \epsilon_{ij}, \quad \epsilon_{ij} \sim \mathcal{N}(0, R), \\
    \lambda_i^{k \to k'}(t_0, t) = \lambda_0^{k \to k'}(t_0, t) \exp\left( \alpha^{k \to k'} g^{k \to k'}(t, \psi_i) + \beta^{k \to k'} X_i \right), \\
    \psi_i = f(\gamma, X_i, b_i), \quad b_i \sim \mathcal{N}(0, Q).
\end{cases}
\]

Here, $h$ and $f$ define the nonlinear mixed-effects submodel,
$g^{k \to k'}$ represents the link between biomarker dynamics and the transition intensity, and $\lambda_0^{k \to k'}$ denotes the baseline hazard associated with transition $(k \to k')$. This structure extends the classical joint model to arbitrary directed graphs and to both Markovian and semi-Markov specifications. 

The coefficient $\alpha^{k \to k'}$ quantifies the effect of the longitudinal biomarker dynamics on the instantaneous risk of transitioning from state $k$ to state $k'$. More concretely, for a one-unit increase in $g^{k \to k'}(t,X_i,\psi_i)$ (holding other covariates constant), the hazard ratio for transition $(k \to k')$ is $\exp(\alpha^{k \to k'})$. Likewise, $\beta^{k \to k'}$ represents the effect of baseline (or time-varying) covariates $X_i$ on that same transition: a one-unit increase in a covariate $X_{ij}$ multiplies the hazard by $\exp(\beta^{k \to k'}_j)$. Thus, $\alpha^{k \to k'}$ captures the degree to which the biomarker trajectory (via its latent parameters $\psi_i$ or function $g^{k \to k'}$) influences the transition risk, while $\beta^{k \to k'}$ captures the direct effect of covariates on that transition’s risk, beyond the biomarker pathway.

\subsubsection{Model Variants}

Two baseline hazard conventions are common and can be specified as
\[
    \lambda_0^{k \to k'}(t_0, t) =
    \begin{cases}
        \lambda_0^{k \to k'}(t - t_0) &\text{(clock-reset)},   \\
        \lambda_0^{k \to k'}(t)  &\text{(clock-forward)}.
    \end{cases}
\]

The \emph{clock-reset} form models risk as a function of time spent in the current state, whereas the \emph{clock-forward} form measures risk with respect to global time since study entry. Common parametric baseline hazard specifications include the exponential ($\lambda_0(t) = \lambda$), the Weibull ($\lambda_0(t) = k \lambda^k t^{k-1}$), and the Gompertz ($\lambda_0(t)= a e^{bt}$) families, among others. 

Furthermore, under the sojourn-time formulation, transition times are no longer constrained to be positive, since only the elapsed duration between successive transitions enters the baseline hazard function.

\subsubsection{Likelihood Formulation}

To derive the marginal likelihood, we decompose the joint distribution of the longitudinal and multi-state processes under a set of standard conditional independence assumptions detailed in Table~\ref{ass:likelihood} below.

\begin{tcolorbox}[title=\textbf{Assumptions for Likelihood Factorization}, colback=gray!4!white,colframe=gray!60!black,sharp corners,boxrule=0.5pt] \label{ass:likelihood}
    \textbf{A. Latent-level independence}
    \vspace{-1em}
    \begin{enumerate}[label=\textbf{A\arabic*.}, leftmargin=3em]
        \item Random effects $(b_i)_i$ are mutually independent across individuals.
    \end{enumerate}
    \vspace{-0.5em}
    \textbf{B. Conditional independence within individuals}
    \vspace{-1em}
    \begin{enumerate}[label=\textbf{B\arabic*.}, leftmargin=3em]
        \item Longitudinal observations $(Y_{ij})_{ij}$ are mutually independent given $b_i, X_i$.
        \item Trajectories $(\mathcal{T}_i^*)_i$ are mutually independent given $b_i, X_i$.
        \item The longitudinal and event processes are mutually independent given $b_i, X_i$.
    \end{enumerate}
    \vspace{-0.5em}
    \textbf{C. Censoring and process assumptions}
    \vspace{-1em}
    \begin{enumerate}[label=\textbf{C\arabic*.}, leftmargin=3em]
        \item[\textbf{C1.}] Censoring times $(C_i)_i$ are mutually independent and noninformative given $b_i, X_i$, \textit{i.e.}, $C_i \indep (Y_i, \mathcal{T}_i^*) \mid b_i, X_i$.
        \item[\textbf{C2.}] Event trajectories satisfy the semi-Markov property~\ref{ass:semi-markov} given $b_i, X_i$.
    \end{enumerate}
\end{tcolorbox}

These grouped assumptions mirror those used in classical joint modeling \citep{rizopoulos2012joint} and multi-state survival analysis \citep{putter2007competing}: independence across subjects, the conditional independence structure linking the longitudinal and event submodels through shared random effects, noninformative censoring, and semi-Markovian dynamics.

\bigbreak

Let $\theta \coloneqq (\gamma, Q, R, \alpha, \beta)$ denote the vector of model parameters. For subject $i \in \llbracket n \rrbracket$, we observe the longitudinal measurements $Y_i = (Y_{i1}, \dots, Y_{in_i})$ and the event trajectory $\mathcal{T}_i = \left\{ (T_{i\ell}, S_{i\ell})_{\ell \leq m_i} \right\}$. The joint likelihood then factorizes as
\[
    p_\theta(Y_i, \mathcal{T}_i \mid X_i, C_i) = \int p_\theta(Y_i \mid b_i, X_i) \, p_\theta(\mathcal{T}_i \mid b_i, X_i, C_i) \, p_\theta(b_i) \, db_i.
\]

This formulation generalizes the joint likelihoods of \citet{wulfsohn1997joint} and \citet{ferrer2016joint} to arbitrary multi-state event structures. Each part can then be explicitly expressed using Assumptions~\ref{ass:likelihood}, yielding an expression very similar to that obtained by \citet{rizopoulos2012joint}.

\begin{description}
    \item[\textnormal{\textbf{Prior likelihood:}}]
        \[
            p_\theta(b_i) = \frac{(2 \pi)^{-q/2}}{\det(Q)^{1/2}}
              \exp\left( -\tfrac{1}{2} b_i^T Q^{-1} b_i \right).
        \]

    \item[\textnormal{\textbf{Longitudinal likelihood:}}]
        \begin{align*}
           &p_\theta(Y_i \mid b_i, X_i) = \prod_{j = 1}^{n_i} \frac{(2 \pi)^{-d/2}}{\det(R)^{1/2}} \exp\left( -\tfrac{1}{2} \left(Y_{ij} - h(t_{ij}, \psi_i) \right)^T R^{-1} \left( Y_{ij} - h(t_{ij}, \psi_i) \right) \right), \\
           &\text{with } \psi_i = f(\gamma, X_i, b_i).
        \end{align*}

    \item[\textnormal{\textbf{Semi-Markov likelihood:}}]
          \begin{equation}
              \begin{aligned}
                  p_\theta(\mathcal{T}_i \mid b_i, X_i, C_i) = &\prod_{\ell=0}^{m_i-1} p_\theta\left((T_{i(\ell+1)}, S_{i(\ell+1)}) \mid b_i, X_i, (T_{i\ell}, S_{i\ell}) \right) \\
     &\exp\left( -\sum_{j: (S_{im_i}, j) \in E} \Lambda_{i, \theta}^{S_{im_i} \to j} (T_{im_i}, C_i \mid b_i, X_i) \right).
              \end{aligned}
              \label{eq:semi-markov}
          \end{equation}
\end{description}

The proof of the expression of the Semi-Markov likelihood
(\ref{eq:semi-markov}) is provided in Appendix
\ref{sec:proof-semi-markov}. Each term $p_\theta\left((T_{i(\ell+1)}, S_{i(\ell+1)}) \mid b_i, X_i, (T_{i\ell}, S_{i\ell}) \right)$ can be computed using Equation~\ref{eq:transition-density}.

\bigbreak

The likelihood factorization above is valid and forms the basis for scalable inference procedures described in Section~\ref{sec:inference}, leveraging the complete trajectory of biomarkers to refine predictions of transitions and survival \citep{de2010mstate}.

Moreover, here, we assume that the initial state $S_{i0}$ is observed. However, the framework could also incorporate a multinomial model for unobserved initial states \citep{yiu2018clustered}.

\section{Statistical Inference and Model Selection} \label{sec:inference}

In this section, we outline a practical framework for statistical inference in the proposed multi-state semi-Markov joint model. We first derive an optimization scheme for estimating model parameters through a stochastic gradient ascent procedure, and then introduce model selection criteria to assess competing specifications. These methods will be illustrated and evaluated on simulated and real datasets in Section~\ref{sec:experiments}.

\subsection{Stochastic Gradient Ascent}

The estimation of model parameters $\theta = (\gamma, Q, R, \alpha, \beta)$ relies on two likelihood formulations: the complete-data likelihood $\mathcal{L}_{\mathrm{complete}}$, which includes latent variables, and the marginal likelihood $\mathcal{L}_\mathrm{marginal}$, obtained by integrating them out
\[
    \mathcal{L}_\mathrm{marginal}(\theta ; X, Y, \mathcal{T}, C) = \int \mathcal{L}_\mathrm{complete}(\theta; X, Y, \mathcal{T}, C, b) \,db,
\]
where $b \coloneqq (b_1, \dots, b_n)$, and owing to the independence across individuals
\[
    \mathcal{L}_\mathrm{complete}(\theta  ; X, Y, \mathcal{T}, C, b) = \prod_{i=1}^n p_\theta(Y_i \mid b_i, X_i) \, p_\theta(\mathcal{T}_i \mid b_i, X_i, C_i) \, p_\theta(b_i).
\]

It is to be noted that this integral is typically intractable in closed form due to the nonlinear nature of the model. Several optimization methods adapted from the nonlinear mixed-effects literature can be applied, including Stochastic EM \citep{kuhn2004coupling}, Laplace approximation \citep{wolfinger1993laplace} and Gauss--Hermite quadrature, alongside MCMC-based approximations such as Metropolis--Hastings \citep{hastings1970monte}, Metropolis-within-Gibbs, Hamiltonian Monte Carlo \citep{neal2011mcmc}. These strategies are implemented in software such as \texttt{JMBayes} \citep{jmbayes22024,rizopoulos2020package}.

\bigbreak

Another approach requiring mild regularity assumptions on the $\log$ marginal likelihood, but without the need for the model to belong in the exponential family, is to consider a stochastic gradient ascent scheme \citep{caillebotte2025estimation, baey2023efficient} using Fisher's identity and following the Robbins-Monro procedure \citep{robbins1951stochastic}.

Under interchangeability of integration and differentiation, setting $x \coloneqq (X, Y, \mathcal{T}, C)$ for convenience, the Fisher identity writes
\[
    \nabla_\theta \log \mathcal{L}_\mathrm{marginal}(\theta; x) = \mathbb{E}_{b \sim p_\theta(\cdot \mid x)} \left[ \nabla_\theta \log \mathcal{L}_{\mathrm{complete}}(\theta; x, b) \right].
\]

This expectation is approximated using Monte Carlo samples from the posterior $p_\theta(\cdot \mid x)$, avoiding the need to evaluate the intractable marginal likelihood $\mathcal{L}_\mathrm{marginal}(\theta; x)$. To ensure the covariance matrices $Q$ and $R$ remain symmetric and positive-definite during optimization, we parametrize them using a $\log$-Cholesky decomposition of their inverse. By taking the logarithm of the diagonal elements of the Cholesky factors, we map the constrained covariance parameters to an unconstrained real-valued space, and avoid explicit matrix inversion for likelihood computation. This formulation naturally supports stochastic gradient ascent with Markov Chain Monte Carlo (MCMC)-based posterior sampling and enables parallelization across individuals, ensuring convergence to a critical point. The update rule is detailed in Algorithm~\ref{alg:sgd}.

\begin{algorithm}[H]
    \caption{Stochastic gradient ascent model inference}
    \label{alg:sgd}

    \begin{algorithmic}[1]
        \algrenewcommand\algorithmicrequire{\textbf{Input:}}
        \Require data $x$; initial parameters $\theta^{(0)}$; step sizes $(\eta_t)_{t \geq 0}$ with $\sum_{t \geq 0} \eta_t = +\infty$, $\sum_{t \geq 0} \eta_t^2 < +\infty$; number of parallel chains $K$; MCMC samplers $\{ p_k^{\mathrm{MCMC}} \}_{k=1}^K$; stopping criterion.
        \Statex
        \State $t \gets 0$
        \While{not converged}
        \State For each $k \in \llbracket K \rrbracket$, draw $b_k \sim p_{\theta^{(t)}}(\cdot \mid x) \approx p_k^\mathrm{MCMC}$.
        \State $\theta^{(t+1)} \gets \theta^{(t)} + \frac{\eta_t}{K} \sum_{k=1}^K \nabla_\theta \log \mathcal{L}_\mathrm{complete} (\theta^{(t)} ; x, b_k)$
        \State $t \gets t + 1$
        \EndWhile
        \State \Return $\hat{\theta}$
    \end{algorithmic}
\end{algorithm}

Preconditioning matrices may also be used, such as the Fisher Information Matrix in a natural gradient ascent framework, or even specialized \emph{optimizers} such as Adam \citep{kingma2014adam}, NAdam \citep{dozat2016incorporating} or Adagrad \citep{duchi2011adaptive}. In particular, the Adam optimizer will later be used in our numerical experiments. Even if its properties do not meet our criteria for convergence, Adam still exhibits stable and smooth optimization trajectories, and is widely used in practical applications.

\subsection{Model Selection and Information Criteria} \label{subsec:criteria}

To compare competing model specifications, one typically relies on information criteria based on the marginal likelihood. In practice, competing models are compared by selecting the specification that yields the smallest value of the chosen information criterion. For instance, the Akaike Information Criterion (AIC) \citep{akaike2003new} and the Bayesian Information Criterion (BIC) \citep{schwarz1978estimating} are standard tools. While the AIC relies on an asymptotically unbiased estimator of the $\log$-likelihood, the BIC aims to identify the true model with high probability as the sample size grows.

The Akaike Information Criterion is computed as follows
\[
    \mathrm{AIC} \coloneqq -2 \log \mathcal{L}_\mathrm{marginal}(\hat{\theta}; X, Y, \mathcal{T}, C) + 2 \dim \hat{\theta},
\]
where $\dim \hat{\theta}$ is the number of parameters, and $\hat{\theta}$ is a maximizer of the marginal likelihood.

The Bayesian Information Criterion is similarly defined but imposes a stronger penalty for model complexity, growing in $\log n$ where $n$ is the number of observations. Particularly in the context of mixed effects models, and even more so for multi-state joint models, the number of observations $n$ may vary from one definition to another, either the total number of repeated measurements or the number of individuals \citep{delattre2014note}. To alleviate this problem, we recall the derivation of the BIC from the Laplace approximation \citep{lebarbier2006introduction}. For a prior distribution $\pi$ on a set of models $\mathcal{M}$, the Laplace approximation yields, up to constant terms (which could also be included to improve the approximation), the approximation of the posterior probability given observed data $x$ reads
\[
    \forall m \in \mathcal{M}, \, \log p(x \mid m) \approx \log \mathcal{L}_\mathrm{marginal}(\hat{\theta}; x) - \frac{1}{2} \log \det(-H_{\hat{\theta}}),
\]
where $H_{\hat{\theta}}$ denotes the Hessian matrix of the marginal $\log$-likelihood evaluated at the Maximum Likelihood Estimator (MLE). Then, under standard regularity conditions that allow differentiation and integration to be interchanged twice, we have that
\[
    -\frac{1}{n} H_{\hat{\theta}} \overset{\mathbb{P}}{\to} \mathcal{I}(\hat{\theta}),
\]
where $\mathcal{I}(\hat{\theta})$ denotes the Fisher Information Matrix \citep{casella2001theory}. Given a reliable estimate $\widehat{\mathcal{I}}(\hat{\theta})$ of the Fisher Information Matrix (see, e.g., \citet{delattre2023computing}), we can then approximate the BIC by substituting this estimate
\[
    \forall m \in \mathcal{M}, \, \log p(x \mid m) \approx \log \mathcal{L}_\mathrm{marginal}(\hat{\theta}; x) - \frac{1}{2} \log \det \hat{\mathcal{I}}_n(\hat{\theta}),
\]
where $\hat{\mathcal{I}}_n(\hat{\theta}) \coloneqq n \, \hat{\mathcal{I}}(\hat{\theta})$, which in practice corresponds to the matrix readily computed by many software packages. 

Thus, our BIC criterion can be written as
\[
\mathrm{BIC}_\mathcal{I} \coloneqq -2 \log \mathcal{L}_\mathrm{marginal}(\hat{\theta}; x) + \log \det \hat{\mathcal{I}}_n(\hat{\theta}).
\]

However, as previously noted, the marginal $\log$-likelihood
\[
\log \mathcal{L}_\mathrm{marginal}(\theta; x)
= \log \int \mathcal{L}_\mathrm{complete}(\theta; x, b) \, db
\]
is generally intractable in mixed effects and joint models, since the integral with respect to the latent variables does not admit a closed form. Although the gradient of the marginal $\log$-likelihood can be efficiently estimated using Fisher's identity, the marginal $\log$-likelihood itself cannot be evaluated directly.

One possibility is to estimate it using bridge sampling, which has been proven to provide consistent and accurate estimators of normalizing constants and marginal likelihoods based on MCMC output \citep{meng1996simulating, gronau2017tutorial}. 

\bigbreak

An alternative is to approximate the posterior distribution $p_{\hat{\theta}}(b \mid x)$ by a Gaussian distribution, for instance via a Laplace approximation. Using the entropy identity,
\[
\log \mathcal{L}_\mathrm{marginal}(\hat{\theta}; x) = \mathbb{E}_{b \sim p_{\hat{\theta}}(\cdot \mid x)} \left[ \log \mathcal{L}_\mathrm{complete}(\hat{\theta}; x, b) \right] - \mathbb{E}_{b \sim p_{\hat{\theta}}(\cdot \mid x)} \left[ \log p_{\hat{\theta}}(b \mid x) \right],
\]
we can further leverage independence across individuals, so that the entropy term decomposes as
\[
\mathbb{E}_{b \sim p_{\hat{\theta}}(\cdot \mid x)} \left[ \log p_{\hat{\theta}}(b \mid x) \right] = \sum_{i=1}^n \mathbb{E}_{b_i \sim p_{\hat{\theta}}(\cdot \mid x_i)} \left[ \log p_{\hat{\theta}}(b_i \mid x_i) \right].
\]

Approximating each posterior $p_{\hat{\theta}}(\cdot \mid x_i)$ by a Gaussian distribution with covariance matrix $\Sigma_i$, we obtain
\[
\log \mathcal{L}_\mathrm{marginal}(\hat{\theta}; x) \approx \mathbb{E}_{b \sim p_{\hat{\theta}}(\cdot \mid x)} \left[ \log \mathcal{L}_\mathrm{complete}(\hat{\theta}; x, b) \right] + \frac{1}{2} \sum_{i=1}^n \left\{ \log \det \Sigma_i + q\left( \log(2\pi) + 1 \right) \right\},
\]
where $q$ denotes the dimension of the latent vector.

This approximation replaces the intractable posterior entropy with a sum of closed-form entropies of $q$-dimensional Gaussian distributions. Furthermore, each covariance matrix $\Sigma_i$ can be easily estimated from MCMC samples of the latent variables using the empirical covariance estimator. The practical relevance of the proposed model selection criteria is discussed in Subsection~\ref{subsec:sim-selection}.

\section{Dynamic prediction} \label{sec:prediction}

In this section, we turn to dynamic prediction within the multi-state framework. Although the prediction of longitudinal trajectories does not fundamentally differ from the standard nonlinear joint modeling setting, the key difference arises from the structure of the event process. Indeed, because the joint distribution of multi-state trajectories has no tractable closed-form expression, direct calculation of predictive quantities is generally infeasible.

To address this, we rely on simulation-based procedures which are rendered possible using Algorithm~\ref{alg:trajectory-sim} (detailed in the Appendix). Specifically, predictions are derived from simulated event trajectories, allowing for individualized risk assessment and forecasting of future states or event times based on a subject's observed biomarker and event history. This approach generalizes classical dynamic prediction by leveraging the rich structure of multi-state processes and the joint distribution of longitudinal and event data.

\bigbreak

Let $i$ index a new individual. Suppose we are interested in some functional of the true (unobserved) future trajectory,
\[
    \chi^*(\mathcal{T}_i^*) \in \mathbb{Y},
\]
where $\mathbb{Y}$ is some generic vector space. 

For example, one can think of the state taken by the individual $i$ at a time $u$. As is the case in traditional dynamic prediction however, we are only interested in predicting quantities or characteristics that depend on a yet unobserved \emph{future}, given prior information, \textit{i.e.}, the trajectory $\mathcal{S}_i(t)$ up to some prediction time $t \leq u$ and observed longitudinal markers $\mathcal{Y}_i(t) \coloneqq \left\{ (t_{ij}, Y_{ij}) : t_{ij} \leq t \right\}$. In contrast to (single transition) joint models, the conditional probability distribution on $(\mathbb{R}^+ \times V)^\mathbb{N}$ cannot be analytically derived.

Nonetheless, as seen in Algorithm~\ref{alg:trajectory-sim}, we are able to accurately simulate each trajectory. Therefore, a single or double Monte Carlo estimation scheme may be devised under certain restrictions, where we estimate $\chi^*(\mathcal{T}_i^*)$ by its conditional expected value (or other statistics depending on the posterior distribution, such as the median)
\[
    \hat{\chi}_i^{(1)} \coloneqq \mathbb{E}_{\hat{\theta}} \left[ \chi^*(\mathcal{T}'_i) \mid \mathcal{Y}_i(t), \mathcal{S}_i(t) \right] \approx \frac{1}{B} \sum_{k=1}^B \chi^*(\mathcal{T}_i^{(k)}), \, \mathcal{T}_i^{(k)} \overset{\mathrm{i.i.d.}}{\sim} \mathscr{L}_{\hat{\theta}}\left( \mathcal{T}_i^* \mid \mathcal{Y}_i(t), \mathcal{S}_i(t) \right),
\]
where the samples $\{ \mathcal{T}_i^{(k)} \}_{k = 1}^B$ are generated using an MCMC algorithm targeting the distribution $\mathscr{L}_{\hat{\theta}}\left( b_i \mid \mathcal{Y}_i(t), \mathcal{S}_i(t) \right)$ in conjunction with Algorithm~\ref{alg:trajectory-sim}.

\bigbreak

Another approach from a Bayesian perspective treats the model parameters $\theta$ as random variables by imposing a prior distribution together with the training data $\mathcal{D}$ \citep{rizopoulos2011dynamic}, and thus in this case our new estimator may be defined
\[
    \hat{\chi}_i^{(2)} \coloneqq \mathbb{E}_{\theta' \sim \mathscr{L}(\theta \mid \mathcal{D})}\left[ \mathbb{E}_{\theta'}\left[ \chi^*(\mathcal{T}'_i) \mid \mathcal{Y}_i(t), \mathcal{S}_i(t) \right] \right],
\]
and approximated by a double Monte Carlo scheme accordingly where $\mathscr{L}(\theta \mid \mathcal{D})$ is the posterior distribution of $\theta$.

\bigbreak

In the case of longitudinal values, the procedure simplifies: for each draw of $b_i$, one usually predicts a single deterministic functional by substituting $\chi$ with $h(\cdot, \psi_i)$, which already corresponds to the conditional mean of $Y_i$ given $b_i$. However, in general multi-state prediction, $\chi^*$ may very well depend on (countably) infinitely many transitions. As a result, since the proposed estimation relies on simulated samples, we require the quantity to depend only on a finite subset of transitions.

\begin{assumption} \label{ass:simulatable}
    There exists  $\chi: \bigcup_{n \geq 1} (\mathbb{R}^+ \times V)^n \to \mathbb{Y}$ and $\tau_i$ a stopping time for the filtration $\mathcal{F}_{in} \coloneqq \sigma\left( (T_{il}, S_{il})_{\ell \leq n} \right)$ such that
    \[
        \begin{cases}
            \tau_i < +\infty \text{ a.s.}, \\
            \chi^*(\mathcal{T}_i^*) = \chi\left( (T_{il}, S_{il})_{\ell \leq \tau_i} \right)
        \end{cases}
    \]
\end{assumption}

Essentially, given Assumption~\ref{ass:simulatable}, with probability one we are guaranteed that the quantity of interest may be computed in a finite number of simulation steps. Indeed, the prediction algorithm for a single sample may be summarized as in Algorithm~\ref{alg:prediction}.

\begin{algorithm}[H]
    \caption{Prediction algorithm}
    \label{alg:prediction}
    \begin{algorithmic}[1]
        \algrenewcommand\algorithmicrequire{\textbf{Input:}}
        \Require $t \in \mathbb{R}^+$, a prediction time; $\mathcal{Y}_i(t)$ marker history; $\mathcal{T}_i$ trajectory up to time $t$; $\chi$ and stopping time $\tau_i$; $\hat{\theta}$ or $\mathscr{L}(\theta \mid \mathcal{D})$.
        \Statex
        \State $\ell \gets \mathrm{len}(\mathcal{T}_i)$
        \While{$\ell \leq \tau_i$}
        \State Append simulated $(T_{i\ell}, S_{i\ell})$ to $\mathcal{T}_i$
        \State $\ell \gets \ell + 1$
        \EndWhile
        \State \Return $\chi\left( (T_{il}, S_{il})_{\ell \leq \tau_i} \right)$
    \end{algorithmic}
\end{algorithm}

Numerous quantities of interest may be encompassed by this framework. We give multiple examples below. First, for a directed acyclic graph $G = (V, E)$, we define the \emph{depth} of $G$, denoted by $\operatorname{depth}(G)$, as the length of the longest directed path in $G$, that is,
\[
    \operatorname{depth}(G) \coloneqq \max_{(v_0, \dots, v_k)} k,
\]
where $(v_0,\dots, v_k)$ ranges over all directed paths in $G$.

\begin{example}[State at time $u$] \label{ex:state-at-time-u}
    Let $G = (V, E)$ be a finite directed acyclic graph, $u \in \mathbb{R}^+$ be a fixed time, and $\mathbb{Y} = \Delta(V)$ be the simplex of distributions over states. Let $\tau_i = \inf \{ n \in \mathbb{N}: T_{in} \geq u \text{ or } S_{in} \text{ is absorbing} \}$. Clearly, $\tau_i \leq \operatorname{depth}(G) < +\infty$. For each state $k \in V$, let $\mathds{1}_k \in \mathbb{Y}$ denote the vector that is $1$ at entry $k$ and $0$ elsewhere (a one-hot encoding of $k$). Then, we set $\chi_u^*(\mathcal{T}_i^*) = \mathds{1}_{S_{i\tau_i}} \in \mathbb{Y}$ to represent the state occupied by individual $i$ at time $u$. In particular, if $V = \{ 0, 1 \}$ and $E = \left\{ (0, 1) \right\}$, we recover the special case used for survival probability estimation in standard joint models \citep{rizopoulos2011dynamic}.
\end{example}

\bigbreak

\begin{example}[Hitting time] \label{ex:hitting-time}
    Let $G = (V, E)$ be a finite directed acyclic graph, $\mathbb{A} \subseteq V$ a non-empty subset of states, and $\mathbb{Y} = \bar{\mathbb{R}}^+$. For any two non-empty sets $\mathbb{A}, \, \mathbb{B} \subseteq V$, we note $\mathbb{A} \rightsquigarrow \mathbb{B}$ if there exists a path from $\mathbb{A}$ to $\mathbb{B}$ in $G$. Let $\tau_i = \inf \{ n \in \mathbb{N} : S_{in} \in \mathbb{A} \vee \{ S_{in} \} \not \rightsquigarrow \mathbb{A} \}$. Then, $\tau_i \leq \operatorname{depth}(G) < +\infty$ and we set $\chi_{\mathbb{A}}^*(\mathcal{T}_i^*) = T_{i\tau_i} + \mathds{1}_{\{ S_{i\tau_i} \} \not \rightsquigarrow \mathbb{A}} (+\infty)$, which represents the hitting time for the set $\mathbb{A}$.
\end{example}

\bigbreak

In Example~\ref{ex:state-at-time-u}, the stopping time $\tau_i$ captures the step at which the process for individual $i$ reaches or exceeds a given time $u$, or enters an absorbing state. The resulting value therefore represents the state of the individual at that specific time.

In Example~\ref{ex:hitting-time}, the stopping time $\tau_i$ corresponds to the first time the process reaches a target subset of states $\mathbb{A}$, or becomes unable to reach it in the future.
The associated value $\chi_{\mathbb{A}}^*(\mathcal{T}_i^*)$ thus represents the \emph{hitting time} of the set $\mathbb{A}$, which may be finite if $\mathbb{A}$ is reached, or $+\infty$ otherwise.

\bigbreak

Other practical applications may include finding the expected number of edges in some given trajectory, the number of times an individual has returned to a specific state, the expected time between transitions\dots

This simulation-based dynamic prediction framework generalizes classical approaches from single-event joint models to the multi-state context. It enables individualized forecasting of future state occupancy, event risks, sojourn times, and other clinically relevant outcomes, fully exploiting the subject's observed biomarker trajectory and event history. In Section~\ref{sec:paquid}, we demonstrate its application to forecasting dependency trajectories in the PAQUID cohort (see Subsection~\ref{subsec:paquid-pred}).

\section{Simulation Study} \label{sec:experiments}

The purpose of this simulation study is to evaluate the finite-sample performance and convergence properties of the proposed inference algorithm under controlled conditions where the true parameters are known. We simulate data from a three-state semi-Markov process coupled with a nonlinear longitudinal biomarker trajectory, representing a simplified disease progression model. This setup allows us to assess both statistical estimation error (see Table~\ref{tbl:convergence-rmse}) and computational scalability of the stochastic gradient estimation procedure. We further evaluate the model selection performance of the proposed methodology, as outlined in Subsection~\ref{subsec:criteria}. To ensure statistical reliability, we perform $n_\mathrm{runs} = 100$ independent simulations on datasets, comprised of $n = 500$ individuals, generated from the same underlying distribution.

All simulations, estimations, and figures reported in this section were produced using the open-source \texttt{jmstate} Python package version 0.17.2 available on \href{https://pypi.org/project/jmstate/}{PyPI}, which we developed to implement the multi-state joint modeling framework and inference algorithms described in Sections~\ref{sec:model} and~\ref{sec:inference}. Reproducible scripts used to generate the experiments are available in its parent \href{https://github.com/felixlaplante0/jmstate}{GitHub} repository, with minor numerical differences expected across platforms or hardware settings, due to the non-deterministic behaviour of some \texttt{PyTorch} operations.

\subsection{Model Specification}

\begin{figure}[H]
    \centering
    \includegraphics[width=0.4\textwidth]{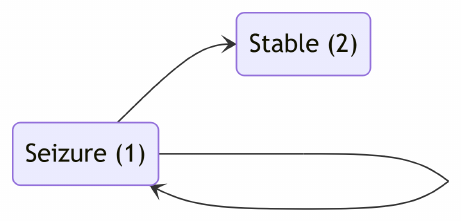}
    \caption{State transition diagram of the simulated recurrent two-state model.}
    \label{fig:sim-graph}
\end{figure}

The simulated model represents a toy example based on a pharmacological scenario where drug absorption can influence seizure activity and patient stability. The model involves two states: Seizure ($1$) and Stable ($2$), which is absorbing. Transitions occur both from $1 \rightarrow 1$ and $1 \rightarrow 2$ (Figure~\ref{fig:sim-graph}), reflecting how a patient may either experience repeated seizure episodes or stabilize after treatment. If the drug is well absorbed, it helps mitigate seizure activity and promotes stabilization. Conversely, poor absorption increases the likelihood of recurrent seizures, highlighting the importance of pharmacokinetics in disease control. Transition intensities follow an exponential baseline hazard.

The longitudinal biomarker trajectory is modeled using a bi-exponential function
\[
    h(t, \psi) = \psi_1 \left( e^{-\psi_2 t} - e^{-\psi_3 t} \right), \quad \psi_i = \gamma e^{b_i}, \quad b_i \sim \mathcal{N} \left( 0, \mathrm{diag}(Q_1, Q_2, Q_3) \right),
\]
and the observation noise is assumed Gaussian, independent across individuals and time points. This bi-exponential structure is commonly used in pharmacokinetic and longitudinal modeling (e.g., \citet{kerioui2022modelling}), although alternative parametrizations are sometimes employed. Incidentally, one of the central points in \citet{kerioui2022modelling} was the deliberate exclusion of recurrent transitions, regarded as difficult to estimate reliably. We show empirically that such transitions can in fact be estimated within our framework.

The link functions are applied consistently across all transitions and defined as
\[
    g^{1 \to 1}(t, \psi) = g^{1 \to 2}(t, \psi) = \int_0^t h(s, \psi) \, ds,
\]
integrating the current biomarker value, thus providing a cumulative measure.

Covariates $X_i$ are normally distributed and enter linearly via $\beta^{k \to k'} X_i$.

Longitudinal data were collected at $20$ equally spaced time points between $t = 0$ and $t = 15$ per subject, with censoring at $C_i \sim \mathcal{U}([10,15])$ alongside the individual trajectories.

\bigbreak

\begin{figure}[H]
    \centering
    \includegraphics[width=0.8\textwidth]{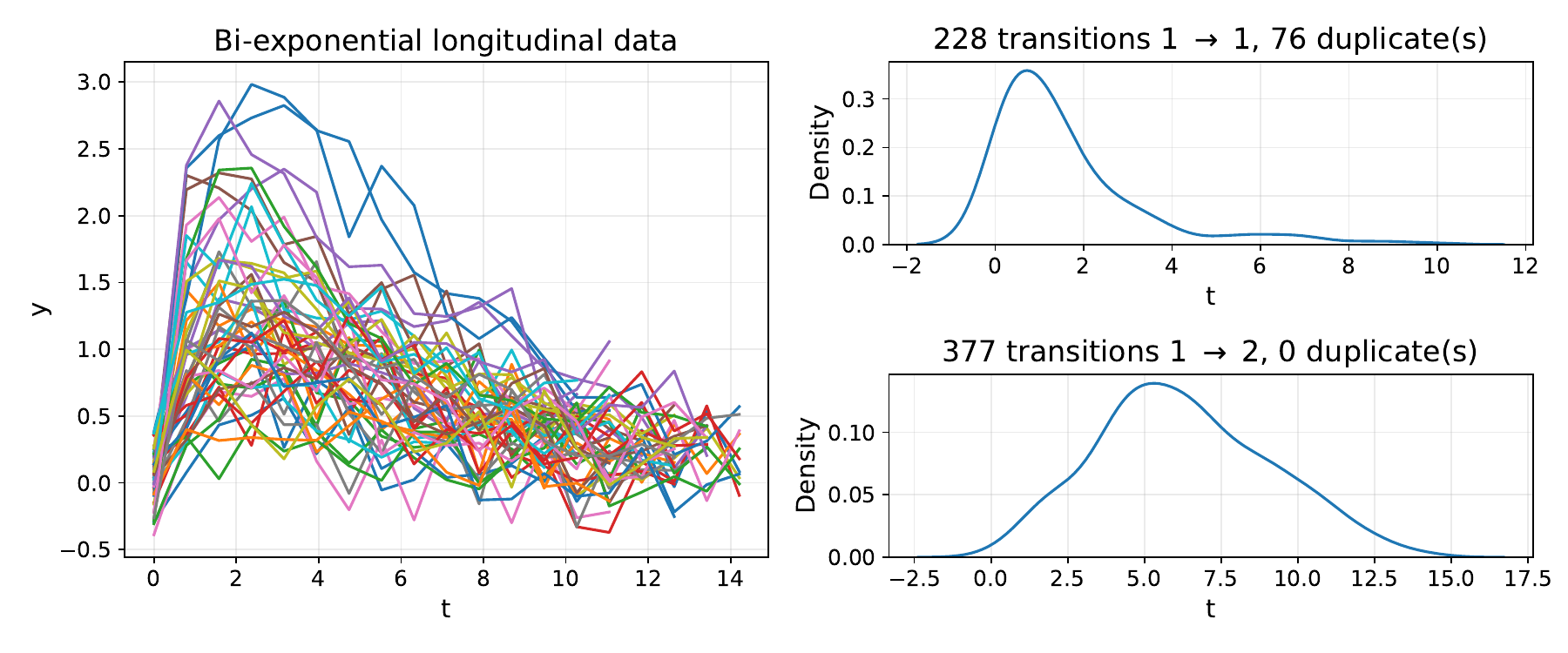}
    \caption{Simulated data: on the left, a sample of longitudinal measurements from $50$ individuals; on the right, observed transitions histograms for the complete population. $76$ patients experienced repeated seizure.}
    \label{fig:simulation}
\end{figure}

Since the base hazard rates $\lambda^{1 \to 1}$ and $\lambda^{1 \to 2}$ are jointly optimized with the other parameters, the final parameter vector lies in $\mathbb{R}^{13}$, and its values used in simulation are shown in Table~\ref{tbl:convergence-rmse}. Each simulated dataset comprised a moderate number of $n = 500$ individuals, generating roughly balanced transition counts (Figure~\ref{fig:simulation}). Thanks to careful parametrization, almost all parameters are mapped to the whole real axis unconstrained.

A short summary of the longitudinal process as well as the trajectories
is given by the Figure~\ref{fig:simulation} below. Note that unlike traditional joint modeling, the notion of censored transitions does not apply here. If a transition does not occur within a trajectory $\mathcal{T}_i$, it is simply omitted from the representation.

\subsection{Estimator Convergence}

For each run, model parameters were estimated using the stochastic gradient ascent procedure described in Section~\ref{sec:inference}, using a Metropolis-within-Gibbs sampling algorithm. Although the initial parameter values can strongly influence the optimization process, we set $\gamma^{(0)} = (1, 1, 1)^T$ to prevent division by zero, initialize both covariance matrices $Q$ and $R$ as identity matrices, set $\lambda^{1 \to 1}$ and $\lambda^{1 \to 2}$ to $1$, and initialize all linear coefficients to zero, thereby avoiding any assumptions about prior knowledge of the parameters.

We used the Adam optimizer alongside an adaptive stopping criterion based on a local linear trend diagnostic applied component-wise to the parameters optimization history. Formally, given a fixed window size $T \geq 2$ and any $t \geq T - 1$, let $ \{ \theta_j^{(s)} \}_{s = t-T+1}^t$ denote the last $T$ iterates of the $j$-th component. We consider the linear model with intercept
\[
    \theta_j^{(s)} = a_j + c_j s + \varepsilon_j^{(s)},
    \qquad s \in \{ t-T+1, \dots, t \},
\]
and compute the associated coefficient of determination $R_{j,t}^2$. Since in such a simple linear regression, the $R^2$ coefficient coincides with the squared correlation between the response and the regressor, small values of $R_{j,t}^2$ indicate that the recent evolution of $\theta_j$ is not better explained by a linear trend than by a constant model, and is therefore consistent with local stationarity.

The optimization procedure is stopped as soon as
\[
    \frac{1}{\dim \theta^{(t)}} \sum_{j=1}^{\dim \theta^{(t)}} R_{j,t}^2 \leq \delta, \label{eq:stop}
\]
where $\delta \in [0, 1]$ is a prescribed threshold (in practice $\delta = 0.1$). This criterion ensures that over the last $T$ iterations, no component displays a statistically meaningful linear drift, so the parameter vector is locally indistinguishable from a constant trajectory. To the best of our knowledge, there is no standard stopping criterion for stochastic optimization procedures; however, in our findings, our method proved to be robust in spite of the stochastic nature of the optimization path, and is scale-invariant. 

\bigbreak

To illustrate the convergence of the estimator, we examine the optimization process for a particular run in Figure~\ref{fig:fitting-plots}, as well as RMSE values computed from $n_\mathrm{runs} = 100$, as shown in Table~\ref{tbl:convergence-rmse}. For a given window size $T = 100$, convergence was typically reached within $500$ stochastic gradient ascent steps, which took an average execution time of around $15$ seconds on an \textit{AMD Ryzen 5 5600} six-core processor. Figure~\ref{fig:fitting-diagnostics} illustrates the stable behaviour of the MCMC sampler over the course of the iterations. We report the mean acceptance ratio and the mean step size for each component of the Metropolis-within-Gibbs sampler, since standard diagnostics such as the effective sample size (ESS) are not directly interpretable in this context because of the non-stationary evolution of the target distribution entailed by the optimization process.

\begin{table}[H]
    \centering
    \begin{tabular}{lcccc}
    \hline
    Parameter &True value &Bias &Std. deviation &RMSE \\
    \hline
    $\gamma_1$ &2.000 &-0.018 &0.174 &0.175 \\
    $\gamma_2$ &0.200 &0.012 &0.120 &0.121 \\
    $\gamma_3$ &1.000 &-0.009 &0.101 &0.101 \\
    $-\frac{1}{2} \log(Q_1)$ &0.949 &-0.017 &0.159 &0.160 \\
    $-\frac{1}{2} \log(Q_2)$ &1.498 &-0.035 &0.237 &0.240 \\
    $-\frac{1}{2} \log(Q_3)$ &1.151 &-0.040 &0.264 &0.267 \\
    $-\frac{1}{2} \log(R)$ &1.498 &0.000 &0.009 &0.009 \\
    $\log \lambda^{1 \to 1}$ &-1.609 &-0.009 &0.152 &0.152 \\
    $\log \lambda^{1 \to 2}$ &-4.605 &-0.035 &0.176 &0.180 \\
    $\alpha_1^{1 \to 1}$ &-1.000 &0.007 &0.148 &0.148 \\
    $\alpha_1^{1 \to 2}$ &0.500 &-0.010 &0.144 &0.144 \\
    $\beta^{1 \to 1}$ &-1.000 &0.025 &0.153 &0.155 \\
    $\beta^{1 \to 2}$ &0.500 &0.018 &0.143 &0.144 \\
    \hline
    \end{tabular}
    \bigbreak
    \caption{Comparison of true and estimated parameters over $100$ runs with $n = 500$ simulated individuals (three decimal places). Since the covariance matrix $Q$ is assumed diagonal, the $\log$-Cholesky reduces to the vector of negative $\log$-standard deviations.}
    \label{tbl:convergence-rmse}
\end{table}

The simulation results indicate that the proposed estimation procedure accurately recovers all parameters of the multi-state joint model for moderate sample sizes. Across the $100$ replications, the biases remain remarkably low, and the RMSE values are small compared to the magnitude of the corresponding parameters. Moreover, the observed variability is probably largely attributable to the intrinsic sampling variability of the true maximum likelihood estimator rather than to optimization error.

The population-level parameters $\gamma$ and the measurement noise parameter $R$ display comparatively lower dispersion, indicating stronger sensitivity for these components within the model. The optimization procedure exhibits stable convergence (Figure~\ref{fig:fitting-plots}) and further supports that the interpretation that residual fluctuations across replications are primarily driven by the variance of the estimator itself.

\subsection{Model Selection} \label{subsec:sim-selection}

To assess the performance of the information criteria described in Subsection~\ref{subsec:criteria}, we compared the Akaike Information Criterion and the modified $\mathrm{BIC}_\mathcal{I}$. Following the same procedure as in the rest of the current section, we performed $n_\mathrm{runs} = 100$ independent simulations, generating for each run a dataset of size $n = 500$ from the above specified model. For each dataset, we then fitted four competing models:

\begin{enumerate}
    \item \textbf{Correct model:} the model used to generate the data, with correctly specified parameters.
    \item \textbf{Shared-$\alpha$ model:} a simplified model in which the link parameters $\alpha^{k \to k'}$ are constrained to be equal across all transitions, \textit{i.e.}, $\alpha^{1 \to 1} = \alpha^{1 \to 2} = \alpha$.
    \item \textbf{Expanded-$\beta$ model:} a slightly more complex model in which three additional normally distributed covariates are added to each individual without any association.  
    \item \textbf{Link misspecified:} a model in which the link functions are incorrectly specified as the bi-exponential function instead of the cumulative exposure.  
\end{enumerate}

For each run, the model with the smallest value of AIC or $\mathrm{BIC}_\mathcal{I}$ was selected. Table~\ref{tbl:model-selection} summarizes the results across the $100$ simulations.  

\begin{table}[H]
    \centering
    \begin{tabular}{lccc}
        \hline
        \textbf{Model} &\textbf{AIC} &$\boldsymbol{\mathrm{BIC}_\mathcal{I}}$ \\
        \hline
        Correct &69 &99 \\
        Expanded-$\beta$ &31 &1 \\
        Shared-$\alpha$ &0 &0 \\
        Misspecified link &0 &0 \\
        \hline
    \end{tabular}
    \bigbreak
    \caption{Number of times each model was selected based on AIC and $\mathrm{BIC}_\mathcal{I}$ over $100$ simulation runs.}
    \label{tbl:model-selection}
\end{table}

The results indicate that both information criteria successfully identify the correct model in the majority of cases. However, $\mathrm{BIC}_\mathcal{I}$ favors the correctly specified model far more frequently than AIC, aligning with its stronger penalty for model complexity, and is able to correctly identify the model $99\%$ of the times. 

Indeed, the increase in the AIC penalty for the expanded model is only $6$ relative to the correctly specified model. Yet the negative log-likelihood is of the order of $2 \, 000$. In such a regime, the additional penalty appears negligible compared with the scale of the likelihood term, so the AIC does not provide a sufficient criterion in this situation. Both the shared-$\alpha$ model and the misspecified link models were never selected by the AIC or the $\mathrm{BIC}_\mathcal{I}$.

Overall, these findings demonstrate that our model selection criteria very effectively distinguish between correctly specified and misspecified multi-state joint models, providing reliable guidance for model choice in finite-sample settings.

\section{Application to the PAQUID Cohort} \label{sec:paquid}

The data used in this section originates from the PAQUID cohort
\citep{Letenneur1994}, a large prospective population-based study initiated in southwestern France in 1988, aimed at understanding the determinants and trajectories of aging. A subsample of $n = 500$ individuals was followed over a period of up to $20$ years \citep{Proust-Lima2017}, with repeated measures of cognitive and physical health, as well as socio-demographic characteristics. 

In particular, global cognitive functioning was assessed through the Mini-Mental State Examination (MMSE), while physical dependency was evaluated using the HIER scale, which classifies subjects into four ordered states (see Figure~\ref{fig:hier-graph}): $0$ (no dependency), $1$ (mild dependency), $2$ (moderate dependency), and $3$ (severe dependency). Individual covariates correspond to a pair of binary variables indicating whether or not the individual has a diploma and is a male $\left( \mathds{1}_\mathrm{CEP}(i), \mathds{1}_\mathrm{male}(i) \right) \in \{0, 1\}^2$.

In this work, we focus on the association between cognitive decline and the progression of physical dependency by jointly modeling the longitudinal trajectory of MMSE scores and the transitions between the four HIER states using a joint multi-state model. More specifically, the state of an individual at a given time $t$ is defined as the highest HIER dependency score between trial entry and $t$, and progression is therefore monotone. This approach allows us to characterize the dynamic interrelationship between cognitive and physical aging processes within a unified statistical framework.

In the following, the full dataset was used to estimate the inferential results presented in Subsection~\ref{subsec:paquid-inference}. For the dynamic prediction analysis in Subsection~\ref{subsec:paquid-pred}, we employed a 5-fold cross-validation procedure, corresponding to splits of $400$ observations for model training and $100$ for validation in each fold, as to obtain unbiased estimates of the predictive performance metrics. The \texttt{jmstate} package was once again employed, and the scripts necessary to reproduce these experiments are also provided in the corresponding repository.

\subsection{Preprocessing}

Longitudinal values were normalized to the interval $[0, 1]$ by dividing each score by its maximum value of $30$, and the measurement times were normalized to represent the elapsed time since age $65$ in decades. This technique ensured stable and fast convergence of the optimization process by keeping all parameters on roughly the same scale.

\subsection{Model Specification}

\begin{figure}[H]
    \centering
    \includegraphics[width=0.9\textwidth]{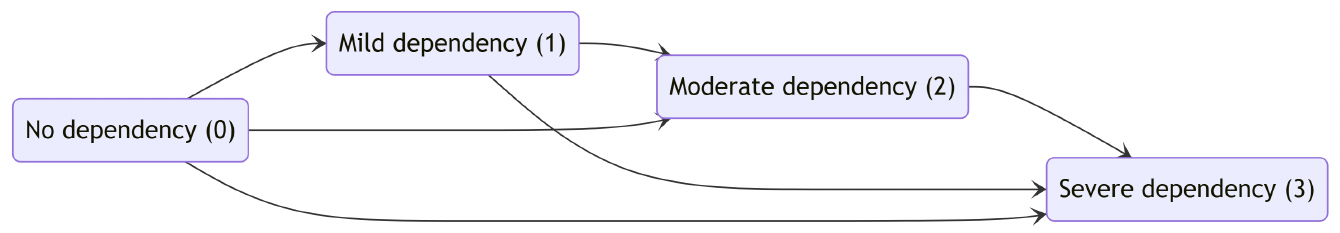}
    \caption{Schematic representation of the four ordered HIER states, with possible transitions. HIER quantifies physical dependency ($0$: no dependency to $3$: severe dependency) \citep{Proust-Lima2017}.}
    \label{fig:hier-graph}
\end{figure}

The transition graph was chosen as follows in Figure~\ref{fig:hier-graph}, and only allows monotonic transitions from a lower level of physical dependency to a higher one, as some transitions occur so infrequently that they cannot be meaningfully analyzed. We considered Weibull baseline hazards for all transitions, consistent with previous observations by \citet{Proust-Lima2017}, with a \emph{clock-reset} (sojourn time) specification. The parameters of these baseline hazards were jointly optimized with the model parameters. It should also be noted that not all patients begin in state $0$, thus, in this respect, our model inherently accounts for left censoring.

The regression and link functions were taken to be shifted and scaled logistic curves such that the normalized MMSE score is always non-increasing
\[
\begin{cases}
    h(t, \psi) = g(t, \psi) = \psi_1 \sigma \left(\frac{\psi_2 - t}{\psi_3} \right),  \quad \sigma(z) \coloneqq \frac{1}{1 + e^{-z}}, \\
    \psi =
    \begin{pmatrix}
        \gamma_1 + \gamma_2 X_1 + \gamma_3 X_2 + b_1   \\
        \gamma_4 + \gamma_5 X_1 + \gamma_6 X_2 + b_2   \\
        \exp(\gamma_7 + \gamma_8 X_1 + \gamma_9 X_2 + b_3)
    \end{pmatrix}.
\end{cases}
\]

For every transition $(k,k') \in E$, we define $g^{k \to k'} := g$. The parameter $\psi_1$ governs the amplitude of the logistic function, $\psi_2$ introduces a time shift, and $\psi_3$ controls the slope at the inflection point.

Moreover, we imposed a specific substructure on the model by setting $\forall (k, k') \in E, \, \beta^{k \to k'} = \beta$, which helped mitigate issues arising from the small number of observations in some groups. 

Additionally, since no prior information was available for the covariance matrix of the latent effects $Q \in \mathbb{R}^{3 \times 3}$, we used a full $\log$-Cholesky parametrization of the precision matrix $Q^{-1}$, with six parameters denoted $\tilde{Q}_1$ through $\tilde{Q}_6$. The full parametrization is essential to capture the potential cross-dependencies between the three latent effects.

\subsection{Inference} \label{subsec:paquid-inference}

The longer optimization time is owed to an effective linear scaling with the number of transitions, as this component of the algorithm cannot be readily vectorized.

\begin{table}[H]
    \centering
    \begin{tabular}{lcc|lcc}
    \hline
    Parameter & Inferred value & Std. error & Parameter & Inferred value & Std. error \\
    \hline
    $\gamma_1$ &1.201 &0.035 &$\boldsymbol{\gamma_2}$ &\textbf{0.221} &\textbf{0.042} \\
    $\boldsymbol{\gamma_3}$ &\textbf{0.155} &\textbf{0.041} &$\gamma_4$ &3.321 &0.076 \\
    $\gamma_5$ &0.181 &0.094 &$\boldsymbol{\gamma_6}$ &\textbf{0.568} &\textbf{0.119} \\
    $\gamma_7$ &0.636 &0.188 &$\boldsymbol{\gamma_8}$ &\textbf{0.663} &\textbf{0.229} \\
    $\boldsymbol{\gamma_9}$ &\textbf{1.168} &\textbf{0.233} &$\tilde{Q}_1$ &2.537 &0.079 \\
    $\tilde{Q}_2$ &-0.032 &0.321 &$\tilde{Q}_3$ &0.644 &0.086 \\
    $\tilde{Q}_4$ &-1.909 &0.134 &$\tilde{Q}_5$ &-0.752 &0.100 \\
    $\tilde{Q}_6$ &-0.690 &0.065 &$-\frac{1}{2}\log(R)$ &3.023 &0.008 \\
    $\log \lambda^{0 \to 1}$ &-1.174 &1.090 &$\log k^{0 \to 1}$ &0.269 &0.125 \\
    $\log \lambda^{0 \to 2}$ &1.160 &0.585 &$\log k^{0 \to 2}$ &0.648 &0.142 \\
    $\log \lambda^{0 \to 3}$ &1.663 &0.769 &$\log k^{0 \to 3}$ &0.876 &0.352 \\
    $\log \lambda^{1 \to 2}$ &1.260 &0.519 &$\log k^{1 \to 2}$ &0.434 &0.097 \\
    $\log \lambda^{1 \to 3}$ &2.744 &0.926 &$\log k^{1 \to 3}$ &0.520 &0.283 \\
    $\log \lambda^{2 \to 3}$ &2.239 &0.346 &$\log k^{2 \to 3}$ &0.509 &0.107 \\
    $\alpha^{0 \to 1}$ &1.254 &1.614 &$\boldsymbol{\alpha^{0 \to 2}}$ &\textbf{-3.844} &\textbf{1.281} \\
    $\boldsymbol{\alpha^{0 \to 3}}$ &\textbf{-8.487} &\textbf{1.362} &$\boldsymbol{\alpha^{1 \to 2}}$ &\textbf{-2.167} &\textbf{0.898} \\
    $\boldsymbol{\alpha^{1 \to 3}}$ &\textbf{-8.393} &\textbf{0.957} &$\boldsymbol{\alpha^{2 \to 3}}$ &\textbf{-5.315} &\textbf{0.516} \\
    $\beta_1$ &-0.159 &0.315 &$\beta_2$ &0.152 &0.233 \\
    \hline
    \end{tabular}
    \bigbreak
    \caption{Inferred parameters and their estimated standard deviations. Bold values mean $p < 0.05$ (only the linear coefficients included).}
    \label{tbl:paquid-fit}
\end{table}

All link parameters $\alpha^{k \to k'}$ for transitions $(k, k') \in E$ are negative and statistically significant, except that from state $0$ to $1$, and the most negative coefficients correspond to transitions reflecting rapid deterioration ($0 \to 3$, $1 \to 3$, and $2 \to 3$). This pattern indicates that the current value of the MMSE score strongly influences the risk of progressing toward physical dependence, and lower cognitive performance is associated with a substantially increased transition intensity, especially for direct or fast declines.

Consistent with other analyses conducted on this dataset, both education level (CEP) and female gender are associated with more favorable cognitive and functional trajectories. Specifically, the inferred values of $\gamma_2$, $\gamma_5$, and $\gamma_8$ indicate that individuals with a primary school certificate (CEP) not only start from a significantly higher baseline level of MMSE, but also experience a delayed and slower progression of dementia. Similarly, the positive values of $\gamma_3$, $\gamma_6$, and $\gamma_9$ suggest the same remarks for women relative to men. Regarding the association coefficient $\beta$, the relatively large standard errors imply that the remaining effects are modest and that the link function already captures most of the relevant information.

\subsection{Multi-State Dynamic Prediction} \label{subsec:paquid-pred}

After fitting the model on each training fold, we performed multi-state dynamic prediction on the corresponding validation fold. For each validation fold $V_k$, with $k \in \llbracket 5 \rrbracket$, predictions were computed using a single Monte Carlo integration scheme as described in Section~\ref{sec:prediction}. We assessed the performance of our model using two metrics: the Brier Skill Score (BSS), which is one of the most standard metrics in this context, as well as the accuracy. More specifically, for each new individual $i$, longitudinal measurements and trajectories were truncated at various landmarks $t$, and predictions were made for future time points $u \geq t$ beyond each truncation. We considered the quantity $\chi_u^*(\mathcal{T}_i^*)$ as defined by Example~\ref{ex:state-at-time-u}, which is estimated using the information available up to $t$, \textit{i.e.}, $\left( \mathcal{Y}_i(t), \mathcal{S}_i(t) \right)$.

\bigbreak

Because the complete trajectories are not fully observed in the test sample and the quantity $\chi_u^*(\mathcal{T}_i^*)$ is not available for $u > C_i$, predictions must account for censoring. For a given landmark time $t$, we obtain a collection of probability estimates
\[
\{ \hat{\chi}_{i, u}(t) \} \subset \Delta\left( \{ 0,1,2,3 \} \right),
\]
each computed from the information available up to $t$. The Brier Skill Score for the fold $V_k$ with $k \in \llbracket 5 \rrbracket$ is then defined by
\[
\mathrm{BSS}_k(t,u) \coloneqq 1 - \frac{\mathrm{BS}_k(t,u)}{\mathrm{BS}_k^{(0)}(u)},
\]
where
\[
\begin{cases}
V_k(t) \coloneqq \{ i \in V_k : C_i > t \}, \\
\mathrm{BS}_k(t, u) \coloneqq \frac{\sum_{i \in V_k(t)} \left\Vert \hat{\chi}_{i, u \wedge C_i}(t) - \chi_{u \wedge C_i}^*(\mathcal{T}_i^*) \right\Vert^2}{\# V_k(t)}, \\
\mathrm{BS}_k^{(0)}(u) \coloneqq \min_a \frac{\sum_{i \in V_k(t)} \left\Vert a - \chi_{u \wedge C_i}^*(\mathcal{T}_i^*) \right\Vert^2}{\# V_k(t)}.
\end{cases}
\]

Hence, values of $\mathrm{BSS}_k(t, u)$ close to $1$ indicate a substantial improvement over the optimal constant prediction.

Similarly, the accuracy at $(t,u)$ on the fold $V_k$ is defined as
\[
\mathrm{Acc}_k(t, u) \coloneqq \frac{\sum_{i \in V_k(t)} \mathds{1}\Bigl\{
    \operatorname*{\arg\max}_j \left( \hat{\chi}_{i, u \wedge C_i}(t) \right)_j = S_i(u \wedge C_i) \Bigr\}}{\# V_k(t)} ,
\]
which corresponds to the proportion of correctly identified states in the non-censored subset.

\bigbreak

\begin{figure}[H]
    \centering
    \includegraphics[width=0.9\textwidth]{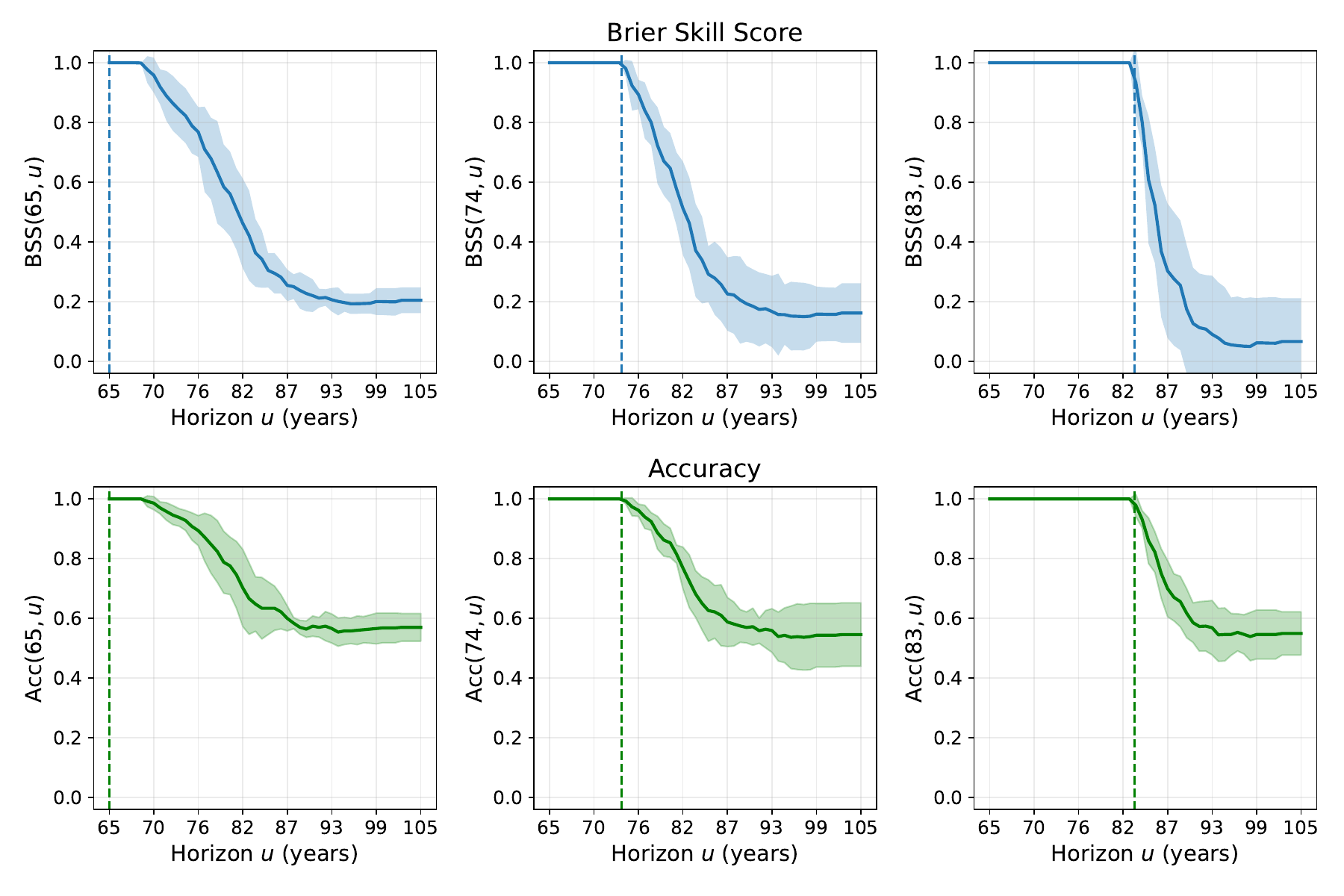}
    \caption{Average values on the $5$-fold cross-validation of the Brier Skill Score (top) and accuracy (bottom) at each time point for different prediction landmarks (dotted lines). Colored bands correspond to $\mathrm{mean} \pm 1.65 \mathrm{sd}$.}
    \label{fig:paquid-acc}
\end{figure}

Figure~\ref{fig:paquid-acc} illustrates our dynamic predictions framework evaluated at three distinct landmark times using both metrics estimated from our cross-validation procedure. Both metrics are averaged across the $5$ folds.

Across the three prediction landmarks, the Brier Skill Score remains above $40\%$ at $10$ years beyond the prediction landmark. This indicates a very strong improvement over the best constant baseline model, despite the limited information available at the time of prediction.

Similarly, the classification accuracy consistently exceeds $60\%$, often higher, thus the model correctly identifies the majority of states. This level of performance persists even for long-term horizons extending more than $20$ years into the future, including during the early phases of the study.

As expected, predictions made immediately following each truncation are the most precise, while uncertainty naturally increases as the prediction horizon extends further. The Brier Skill Score (BSS) drops sharply, which we assume is a byproduct of having very few uncensored participants remaining in the study and the distribution of states quickly converging.

\section{Conclusion} \label{sec:conclusion}

We presented a general likelihood-based framework for joint modeling of nonlinear longitudinal biomarkers and multi-state survival processes. The model unifies both components on arbitrary directed graphs, extending classical joint models to accommodate Markov and semi-Markov structures, recurrent transitions, and nonlinear mixed-effects submodels. This formulation provides a flexible and rigorous foundation for linking longitudinal biomarker trajectories to complex event histories within a single coherent probabilistic model.

Simulation studies confirmed robust parameter recovery, and an application to the PAQUID cohort highlighted the ability to capture the interplay between cognitive decline and physical dependency.

Since the model may potentially involve many transitions, and therefore a large number of parameters, a strategy of parameter sharing across transitions could help reduce the overall number of parameters. This is especially beneficial in settings with limited data for certain transitions, or when biological or clinical knowledge suggests similar effects across multiple transitions. Parameter sharing improves statistical efficiency, reduces overfitting, and facilitates model interpretability.

Future directions include a spline-based parametrization of baseline hazards to allow for nonparametric estimation. This approach is particularly useful when the true baseline hazard is expected to be complex or non-monotonic, or when parametric forms (such as exponential or Weibull) may be too restrictive and lead to model misspecification.

Another promising direction is the development of efficient visualization tools for multi-state trajectories and dynamic predictions. Interactive representations of individual event histories, transition probabilities, and longitudinal biomarker evolution would greatly facilitate model interpretation and communication of results to applied researchers.

\section*{Funding}

This work was partially funded by the Stat4Plant project
ANR-20-CE45-0012.

\section*{Acknowledgments}

The authors would like to thank Jean-Benoist Leger for helpful
discussions.

\subsection*{Conflict of interest}
The authors declare that they have no conflicts of interest.

\subsection*{Author contributions}
F\'elix Laplante: Software, Methodology, Writing – original draft. \\
Christophe Ambroise: Data curation, Writing – original draft.

\subsection*{Data availability}
The data analyzed in this study are publicly available via the R package \texttt{timeROC} (dataset \texttt{Paquid}).

\newpage

\bibliographystyle{sn-basic}
\bibliography{sources}

\appendix

\newpage

\section{Instantaneous Risk Formulae} \label{sec:proof-risk}

For all the following derivations, assume $t \geq t_0$ and that $(k, k') \in E$ is an admissible transition for the graph $G = (V, E)$. As previously stated, we further assume every hazard rate is finite, \textit{i.e.}
\[
\forall (k, k') \in E, \forall t \geq t_0, \, \lambda^{k \to k'}(t_0, t) < +\infty.
\]

\bigbreak

\begin{proof}[\textbf{Proof of Equation~\ref{eq:survival}}]
First, we define the total overall hazard rate of leaving state $k$ at time $t$. Because the process must transition to exactly one of the available next states, we can sum the cause-specific instantaneous risks over all admissible destination states $j$
\[
\sum_{j : (k, j) \in E} \lambda_i^{k \to j}(t_0, t) = \lim_{\delta \to 0^+} \frac{\mathbb{P}(t < T_{i(\ell+1)} \leq t + \delta \mid T_{i(\ell+1)} > t, T_{i\ell}=t_0, S_{i\ell}=k)}{\delta}.
\]

\bigbreak

By the definition of conditional probability, we can rewrite the right-hand side using the survival function $S_i^{k \to \bullet}(t_0, t) \coloneqq \mathbb{P}(T_{i(\ell+1)} > t \mid T_{i\ell}=t_0, S_{i\ell}=k)$. The probability of transitioning in the interval $(t, t+\delta]$ divided by the survival probability up to $t$ gives:
\[
\sum_{j : (k, j) \in E} \lambda_i^{k \to j}(t_0, t) = \frac{1}{S_i^{k \to \bullet}(t_0, t)} \lim_{\delta \to 0^+} \frac{S_i^{k \to \bullet}(t_0, t) - S_i^{k \to \bullet}(t_0, t + \delta)}{\delta} = -\frac{d\log S_i^{k \to \bullet}(t_0, t)}{dt}. 
\]

Next, integrating from the entry time $t_0$ to $t$ and swapping integration and summation, one obtains
\[
\int_{t_0}^t \sum_{j : (k, j) \in E} \lambda_i^{k \to j}(t_0, s) \, ds = \sum_{j : (k, j) \in E} \Lambda_i^{k \to j}(t_0, t) = \underbrace{\log S_i^{k \to \bullet}(t_0, t_0)}_0 - \log S_i^{k \to \bullet}(t_0, t),
\]
since $\mathbb{P}(T_{i(\ell+1)} > t_0 \mid T_{i\ell} = t_0, S_{i\ell} = k) = 1$.

\bigbreak

Taking the exponential of both sides removes the logarithm and directly yields Equation~\ref{eq:survival}.
\end{proof}

\bigbreak

\begin{proof}[\textbf{Proof of Equation~\ref{eq:transition-density}}]
We start with a reformulation of the formal limit definition of the instantaneous risk. We get
\[
\lambda_i^{k \to k'}(t_0, t) = \frac{1}{S_i^{k \to \bullet}(t_0, t)} \lim_{\delta \to 0^+} \frac{\mathbb{P}(t < T_{i(\ell+1)} \leq t + \delta, S_{i(\ell+1)}=k' \mid T_{i\ell}=t_0, S_{i\ell}=k)}{\delta}.
\]

As $\delta \to 0^+$, the right side term evaluates precisely to the joint probability density function of transitioning at time $t$ to state $k'$, which exists since we assumed that every hazard rate is finite, thus
\[
\lambda_i^{k \to k'}(t_0, t) = \frac{p(T_{i(\ell+1)}=t, S_{i(\ell+1)}=k' \mid T_{i\ell}=t_0, S_{i\ell}=k)}{S_i^{k \to \bullet}(t_0, t)},
\]
and multiplying both sides by the survival function, we isolate the joint density
\[
p(T_{i(\ell+1)}=t, S_{i(\ell+1)}=k' \mid T_{i\ell}=t_0, S_{i\ell}=k) = \lambda_i^{k \to k'}(t_0, t) S_i^{k \to \bullet}(t_0, t)
\]

\bigbreak

Finally, substituting the explicit formula for the survival function that we derived in Equation~\ref{eq:survival} provides the final joint density shown in Equation~\ref{eq:transition-density}.
\end{proof}

\bigbreak

\begin{proof}[\textbf{Proof of Equation~\ref{eq:conditional}}]
By Bayes' theorem, the probability of transitioning to state $k'$, given that a transition occurs exactly at time $t$, is the joint density divided by the marginal density of the transition time $T_{i(\ell+1)}$. Using the joint density expression from Equation~\ref{eq:transition-density}, we get
\begin{align*}
\mathbb{P}(S_{i(\ell+1)} = k' \mid T_{i(\ell+1)}=t, T_{i\ell}=t_0, S_{i\ell}=k) &= \frac{\lambda_i^{k \to k'}(t_0, t) S_i^{k \to \bullet}(t_0, t)}{\sum_{j : (k, j) \in E} \left\{ \lambda_i^{k \to j}(t_0, t) S_i^{k \to \bullet}(t_0, t) \right\}} \\
&= \frac{\lambda_i^{k \to k'}(t_0, t)}{\sum_{j : (k, j) \in E} \lambda_i^{k \to j}(t_0, t)}.
\end{align*}

\bigbreak

This leaves only the ratio of the instantaneous risks, resulting exactly in Equation~\ref{eq:conditional}.
\end{proof}

\section{Expression of the Semi-Markov Likelihood~\ref{eq:semi-markov}} \label{sec:proof-semi-markov}

\begin{proof}[\textbf{Proof}]
First, we rely on the fact that, using the semi-Markov property~\ref{ass:semi-markov}
\begin{align*}
    p_\theta\left( (T_{i(\ell+1)}, S_{i(\ell+1)})_{\ell \leq m_i - 1} \mid b_i, X_i \right) &= \prod_{\ell=0}^{m_i - 1} p_\theta\left( (T_{i(\ell+1)}, S_{i(\ell+1)}) \mid b_i, X_i, (T_{i\ell'}, S_{i\ell'})_{\ell' \leq \ell} \right), \\
    &= \prod_{\ell=0}^{m_i - 1} p_\theta\left( (T_{i(\ell+1)}, S_{i(\ell+1)}) \mid b_i, X_i, (T_{i\ell}, S_{i\ell}) \right).
\end{align*}

On the other hand, the probability that no additional event is observed between $T_{im_i}$ and $C_i$ is $1$ if the last state reached is absorbing, and otherwise, since censoring is assumed non-informative, it can be treated as an independent constant
\[
    \mathbb{P}_{\theta}\left( T_{i (m_i+1)} > C_i \mid b_i, X_i, (T_{im_i}, S_{im_i}) \right) = \exp\left(-\sum_{j : (S_{im_i}, j) \in E} \Lambda_{i, \theta}^{S_{im_i} \to j}(T_{im_i}, C_i \mid b_i, X_i) \right),
\]
so we recover the formula cited above, with the convention that an empty sum is $0$.
\end{proof}

\section{Multi-State Simulation}

In nonlinear joint models with known design and known parameters $\theta = (\gamma, Q, R, \alpha, \beta)$, the occurrence times of events conditionally on the latent variables $b_i$ and the covariates $X_i$ can be easily sampled using the inverse cumulative distribution function transform method. This may be achieved through bisection or other root-finding algorithms.

The simulation of the transition process $\mathcal{T}_i^*$ conditionally on $b_i$, $X_i$, and $(T_{i0}, S_{i0})$ can be achieved drawing on the semi-Markov property~\ref{ass:semi-markov} by considering one transition at a time, according to Algorithm~\ref{alg:trajectory-sim}. The algorithm is similar to Gillespie's algorithm \citep{gillespie2007stochastic}, and can also include a survival condition $T_{i1} \geq t_{i, \mathrm{cond}}$ that uses the Chasles relation. 

Here, we deliberately avoid imposing any conditioning in order to preserve maximal generality, since the algorithm is intended to be applicable beyond the specific framework of our model, only assuming continuous transition times distributions and the semi-Markov property.

\begin{algorithm}[H]
    \caption{Simulation of trajectory $i$}
    \label{alg:trajectory-sim}

    \begin{algorithmic}[1]
        \renewcommand{\algorithmicrequire}{\textbf{Input:}}
        \Require subject-specific censoring time $C_i \in \bar{\mathbb{R}}^+$; initial time-state pair $(T_{i0}, S_{i0})$; optional survival condition $T_{i1} \geq t_{i, \mathrm{cond}}$, defaults to $t_{i, \mathrm{cond}} = T_{i0}$.
        \Statex
        \State Initialize $\ell \gets 1$, $\mathcal{T}_i \gets \left( (T_{i0}, S_{i0}) \right)$
        \Statex
        \While{$T_{i(\ell-1)} < C_i$ and $\{ j: (S_{i(\ell-1)}, j) \in E \} \neq \emptyset$}
        \For{each $j$ with $(S_{i(\ell-1)}, j) \in E$}
        \State Draw $T_{i\ell}^{(j)}$ such that $\forall t \geq T_{i(\ell-1)} \vee t_{i, \mathrm{cond}}$
        \Statex \hspace{\algorithmicindent} \hspace{\algorithmicindent} $-\log \mathbb{P}(T_{i\ell}^{(j)} > t) = \int_{T_{i(\ell-1)} \vee t_{i, \mathrm{cond}}}^t \lambda_i^{S_{i(\ell-1)} \to j}(s \mid T_{i(\ell-1)})\, ds$
        \EndFor
        \Statex
        \State Set $T_{i\ell} \gets \min_j T_{i\ell}^{(j)}$, and $S_{i\ell} \gets \arg \min_j T_{i\ell}^{(j)}$
        \State Append $(T_{i\ell}, S_{i\ell})$ to $\mathcal{T}_i$
        \State $\ell \gets \ell + 1$
        \EndWhile
        \Statex
        \If{$T_{i(\ell-1)} > C_i$}
        \State Remove the last pair: $\mathcal{T}_i \gets \mathcal{T}_i[:-1]$
        \EndIf
        \State \Return $\mathcal{T}_i$
    \end{algorithmic}
\end{algorithm}

The proof of the exactness of this algorithm is given below.

\begin{proof}[\textbf{Proof}]
First, consider the case $t_{i, \mathrm{cond}} = T_{i0}$. 

Let $\ell \geq 0$ and $t \geq T_{i\ell}$. For any $k' \in V$ such that $(S_{i\ell}, k') \in E$, calling $J_{i\ell}^{(k')} \coloneqq \{ j : (S_{i\ell}, j) \in E \} \setminus \{ k' \}$, we have
\begin{align*}
    p(T_{i(\ell+1)} = t, S_{i(\ell+1)} = k' \mid T_{i\ell}, S_{i\ell}) &= p\left( \left\{ T_{i(\ell+1)}^{(k')} = t \right\} \cap \left\{ \forall j \in J_{i\ell}^{(k')} \, T_{i(\ell+1)}^{(j)} > t \right\} \mid T_{i\ell}, S_{i\ell} \right) \\
    &\overset{\indep}{=} \lambda_i^{S_{i\ell} \to k'}(T_{i\ell}, t) \begin{aligned}[t] &\exp\left( -\Lambda_i^{S_{i\ell} \to k'}(T_{i\ell}, t) \right) \\ &\exp\left( -\sum_{j \in J_{i\ell}^{(k')}} \Lambda_i^{S_{i\ell} \to j}(T_{i\ell}, t) \right) \end{aligned} \\
    &= \lambda_i^{S_{i\ell} \to k'}(T_{i\ell}, t) \exp\left( -\sum_{j : (S_{i\ell}, j) \in E} \Lambda_i^{S_{i\ell} \to j}(T_{i\ell}, t) \right).
\end{align*}

Thus, we retrieve the same joint density as in Equation~\ref{eq:transition-density}, and right-censoring is then applied accordingly. If $t_{i, \mathrm{cond}} > T_{i0}$, one can also check that both densities are equal by simply integrating from $t_\mathrm{cond}$ instead of $T_{i\ell}$. Note that the conditioning only affects the first transition, as for all subsequent transitions where $\ell > 0$, we have $T_{i\ell} \geq t_{i, \mathrm{cond}} \implies \forall \ell > 0, \, T_{i\ell} \vee t_{i, \mathrm{cond}} = T_{i\ell}$.
\end{proof}

\section{Stochastic optimization process on simulated data}

\vspace{-2em}

\begin{figure}[H]
    \centering
    \includegraphics[width=\textwidth]{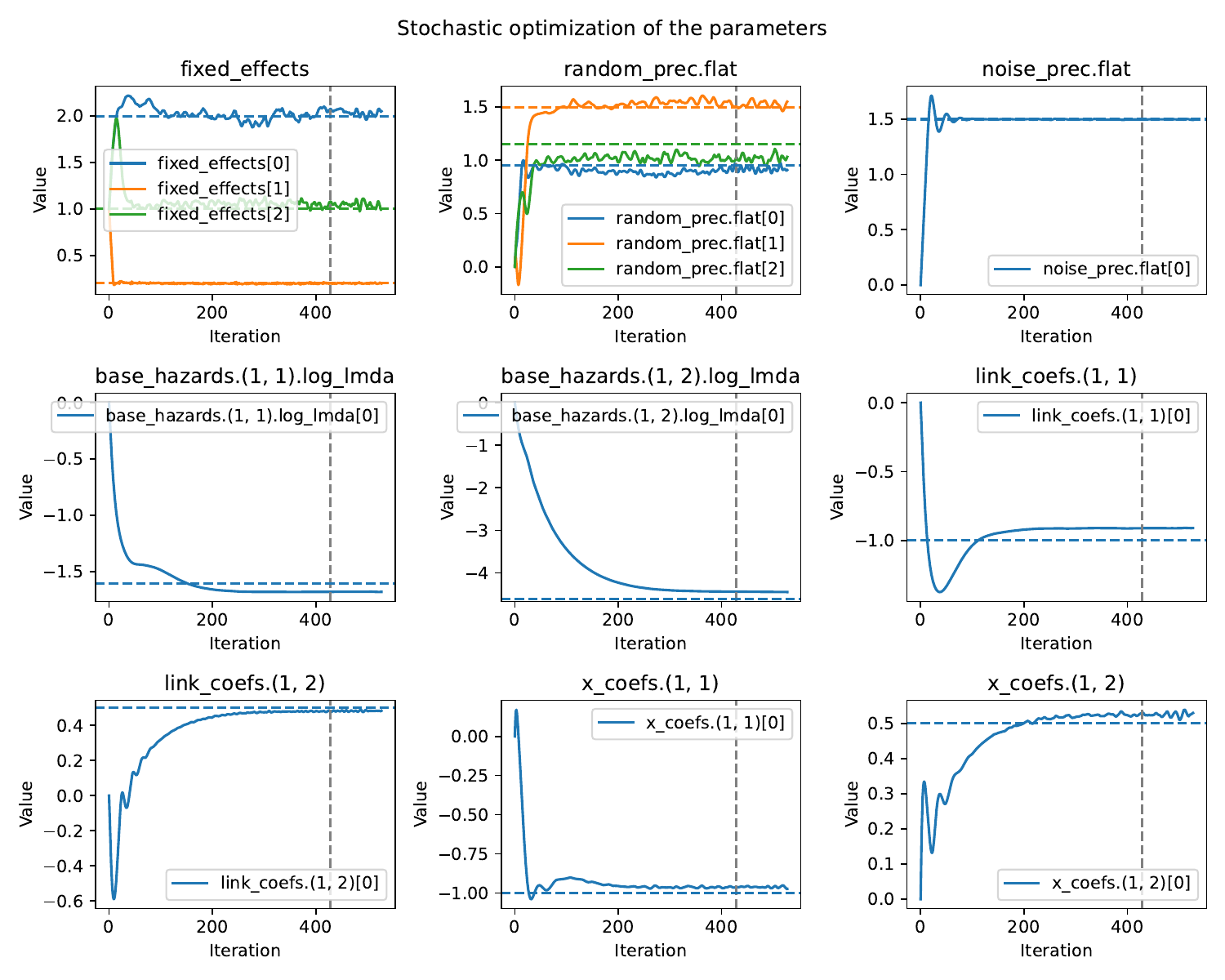}
    \caption{Evolution of the parameters during the optimization of the marginal $\log$-likelihood using stochastic gradient ascent. Dotted horizontal lines correspond to the true values, while vertical lines mark the start of each window used to evaluate the stopping criterion.}
    \label{fig:fitting-plots}
\end{figure}

\vspace{-2em}

\begin{figure}[H]
    \centering
    \includegraphics[width=0.7\textwidth]{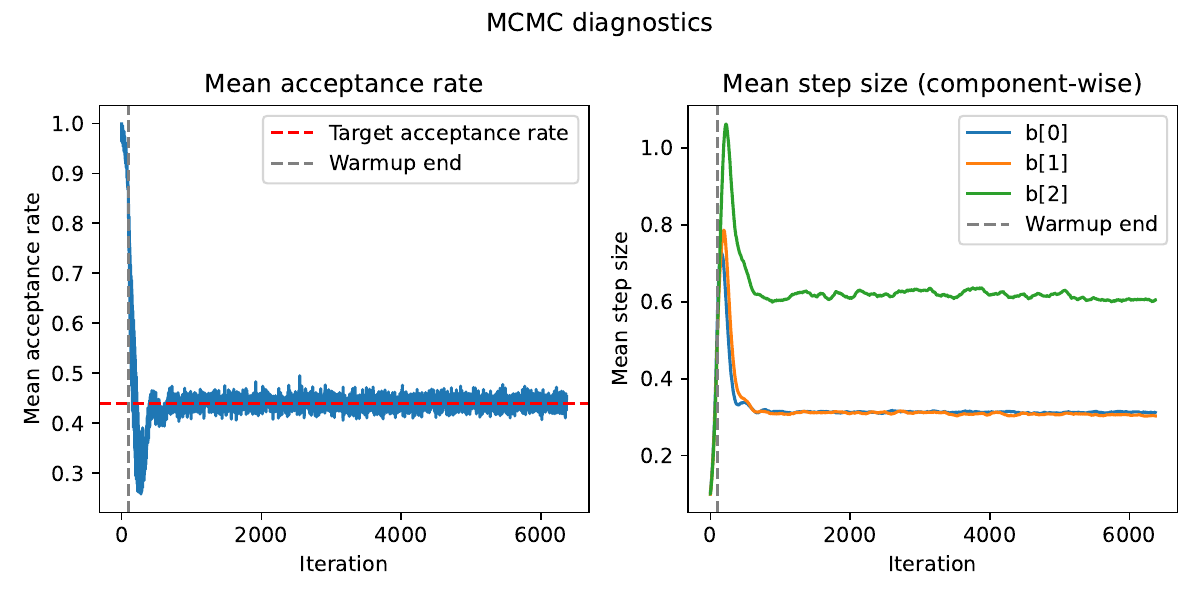}
    \caption{MCMC diagnostics showing mean acceptance rate (left) stabilizing near the target after warmup (gray line), and mean step sizes (right) converging for each parameter component, indicating well-tuned sampling.}
    \label{fig:fitting-diagnostics}
\end{figure}

\section{Stochastic optimization process on the PAQUID dataset}

\vspace{-2em}

\begin{figure}[H]
    \centering
    \includegraphics[width=\textwidth]{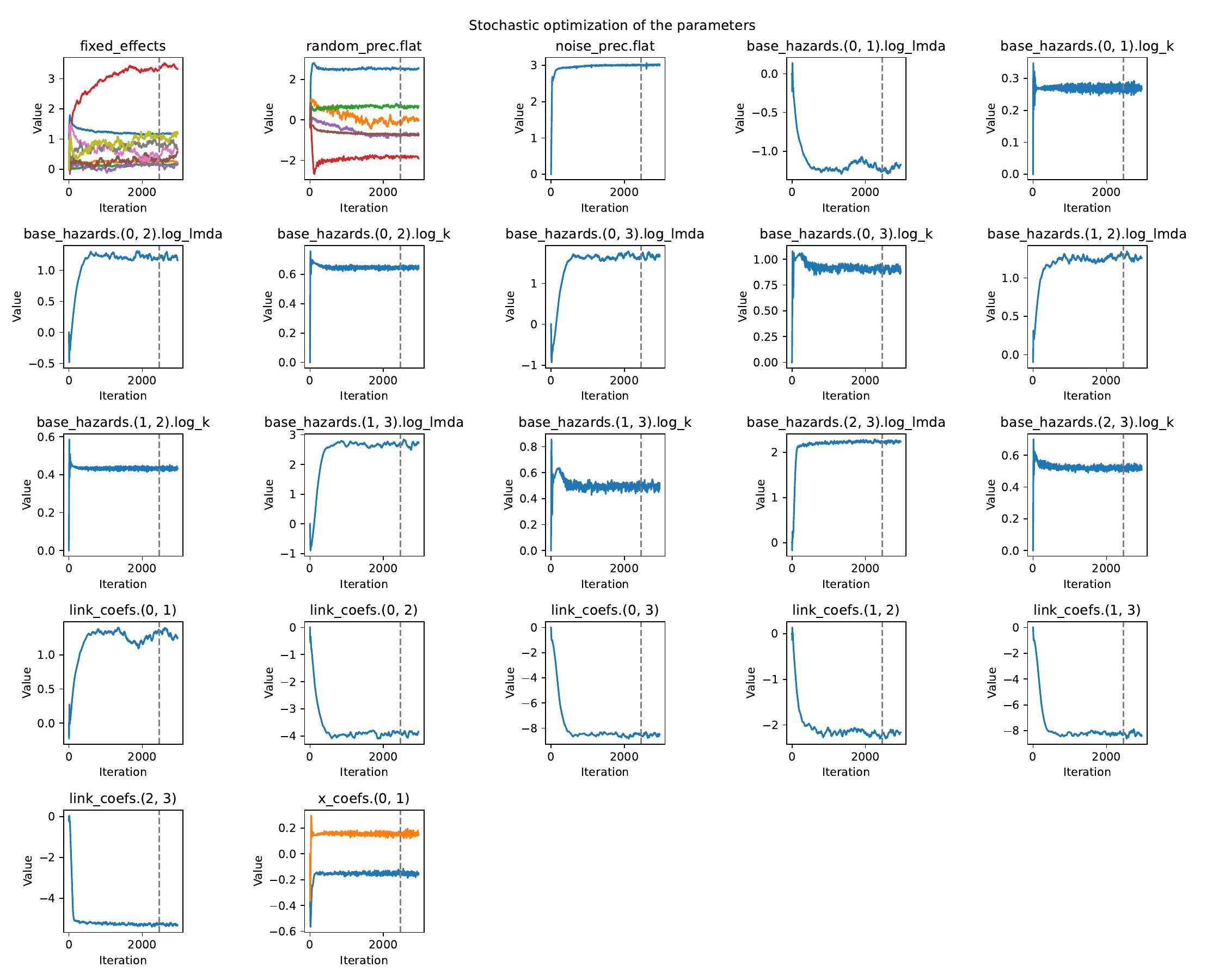}
    \caption{Evolution of the parameters during the optimization of the marginal $\log$-likelihood using stochastic gradient ascent. Vertical lines mark the start of each window used to evaluate the stopping criterion.}
    \label{fig:paquid-plots}
\end{figure}

\vspace{-2em}

\begin{figure}[H]
    \centering
    \includegraphics[width=0.7\textwidth]{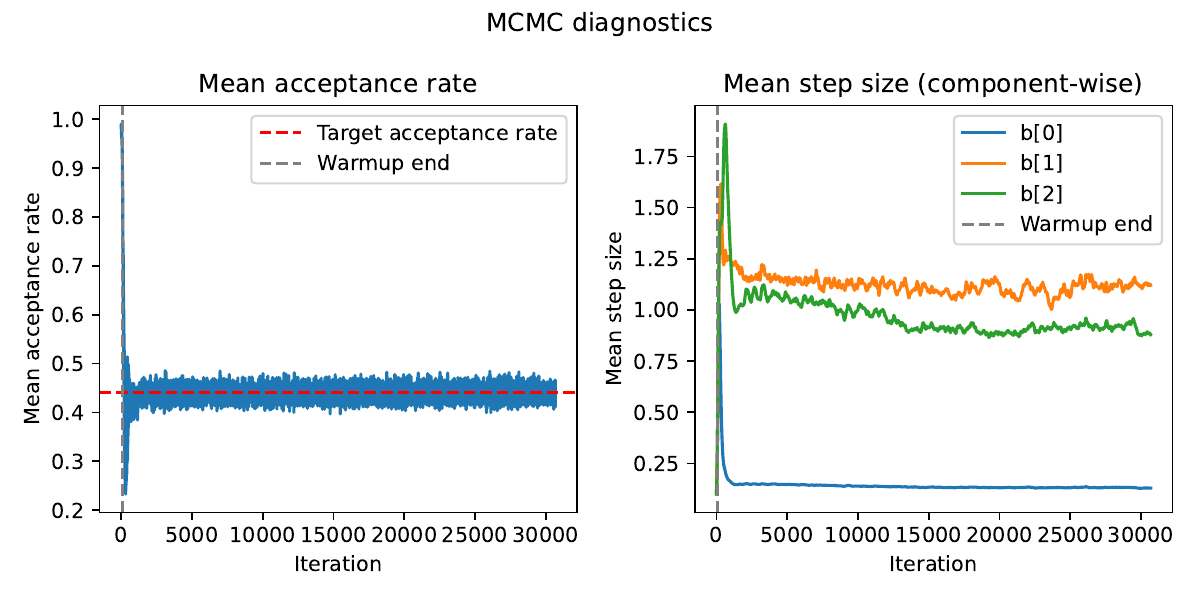}
    \caption{MCMC diagnostics showing mean acceptance rate (left) stabilizing near the target after warmup (gray line), and mean step sizes (right) converging for each parameter component, indicating well-tuned sampling.}
    \label{fig:paquid-diagnostics}
\end{figure}

\end{document}